\documentstyle[11pt,epsf,boxit]{article}
\input epsf         % defines \epsfbox and supporting macros
\epsfverbosetrue    % messages will show height and width
\def\postscript#1#2{\begin{center}\leavevmode \hbox{\epsfxsize=#2 \epsfbox{#1}}\end{center}}

\newcommand{\kk}{{\bf k}}

\def\myfigure#1#2#3{\begin{figure}[t]\begin{boxit} \postscript{#1}{#2\columnwidth}  \caption{\small #3}\label{#1}\end{boxit}\end{figure}}

\title{$a$-$b$ Plane Anisotropy in YBCO
 \thanks{To appear in the proceeding of the {\sc bicas} Summer School
on ``Symmetry of the Order Parameter in High-Temperature
Superconductors.''}}
\author{W. Atkinson, J.P. Carbotte and C. O'Donovan\\
Department of Physics \& Astronomy, McMaster University\\
Hamilton, Ontario, Canada L8S 4M1}
\date{April 15, 1996}
\begin{document}
%initfloatingfigs
\markboth{cond-mat/9604104}{cond-mat/9604104}

\maketitle
\begin{abstract}
The zero temperature in plane penetration depth in an untwinned single
crystal of optimally doped YBa$_2$Cu$_3$O$_{6.93}$ is highly
anisotropic.  This fact has been interpreted as evidence that a large
amount of the condensate resides in the chains.  On the other hand,
the temperature dependence of $\lambda_a(T)/\lambda_a(0)$ and
$\lambda_b(T)/\lambda_b(0)$ (where the $b$-direction is along the
chains) are not very different.  This constrains theories and is, in
particular, difficult to understand within a proximity model with
$d$-wave pairing only in the CuO$_2$ plane and none on the CuO chains
but instead supports a more three dimensional models with interplane
interactions.
\end{abstract}

\section*{Introduction}
\setcounter{page}{1}
\renewcommand{\thepage}{\arabic{page}}

Most theories of the high $T_c$ oxides start with a CuO$_2$ plane
which is the common building block found in all the copper oxide
superconductors.  Such a system is tetragonal and is sometimes
described by simple two dimensional tight binding bands with first and
second nearest neighbour hopping.  The mechanism involved in the
superconductivity is still unknown but there is now strong evidence,
if not yet a consensus, that the gap has d-% wave
symmetry.\cite{olson,shen,ding1,wollman,tsuei1,kirtley,tsuei2} Of
course there are other elements to the structure of the typical copper
oxide superconductor but a stack of CuO$_2$ planes weakly coupled
through a transverse hopping $t_\perp$ can be taken as a simplified
first model.  While in many of the oxides $t_\perp$ is small---perhaps
of the order of 0.1 meV in Bi$_2$Sr$_2$Cu Cu$_2$O$_8$,\cite{zha} and
of the order of a few meV in LaSrCuO$_4$---in
YBa$_2$Cu$_3$O$_{7-\delta}$ (YBCO) at optimum doping, it is much
larger and of the order of a few tens of meV which is almost of the
same order of magnitude as the in-plane hopping integrals and
indicates that this material may be fairly three dimensional.  In
addition, YBCO has chains (CuO) as well as planes (CuO$_2$).  In such
circumstances, the system can be expected to be significantly
orthorhombic with the source of orthorhombicity residing in the
chains.  In an orthorhombic system, the irreducible representation of
the point group crystal lattice which contains the $d_{x^2-y^2}$ part
also contains $s_{x^2+y^2}$ and $s_\circ$ (constant) parts, and these
can mix in the gap so that we cannot expect a pure $d$-wave order
parameter.

%%%%%%%%%%%%%%%%%%%%%%%%%%%%%%%%%%%%%%%%%%%%%%%
% Bill: I moved your \AA out of the equations %
% so that they would be prettier.             %
%%%%%%%%%%%%%%%%%%%%%%%%%%%%%%%%%%%%%%%%%%%%%%%

Strong evidence that the chains play a very important role in
optimally doped YBCO is obtained from infrared and microwave
experiments on untwinned single crystals of YBa$_2$\-Cu$_3$\-O$_{6.93}$.
Far infrared experiments\cite{basov} can be used to measure the
absolute value of the in plane penetration depth at zero temperature
in each of the two principle directions denoted by $a$ and $b$ with
the chains oriented along $b$.  The results are $\lambda_b \sim
1030$\AA\ and $\lambda_a \sim 1600$\AA\ for YBa$_2$Cu$_3$O$_{6.93}$
with $T_c = $93.2K.  On the other hand, the temperature
dependence\cite{bonn,hardy2} of the normalized penetration depth,
$\lambda(T)/\lambda(0)$, is almost the same in both directions.  These
results have been taken as evidence that the chains carry a
significant amount of the condensate density and that the gap in the
chains is of the same order of magnitude as in the planes, a fact
confirmed in current-imaging tunneling spectroscopy ({\sc cits})
experiments.\cite{edwards, edwards2} It would also follow from the
observed large orthorhombicity that the order parameter will not be of
a pure $d_{x^2-y^2}$ symmetry as previously stated and repeated here.
It should contain a significant $s$-admixture due to the existence of
the chains which are coupled to the planes and participate importantly
in the superconductivity.

\section*{Proximity Model}

\myfigure{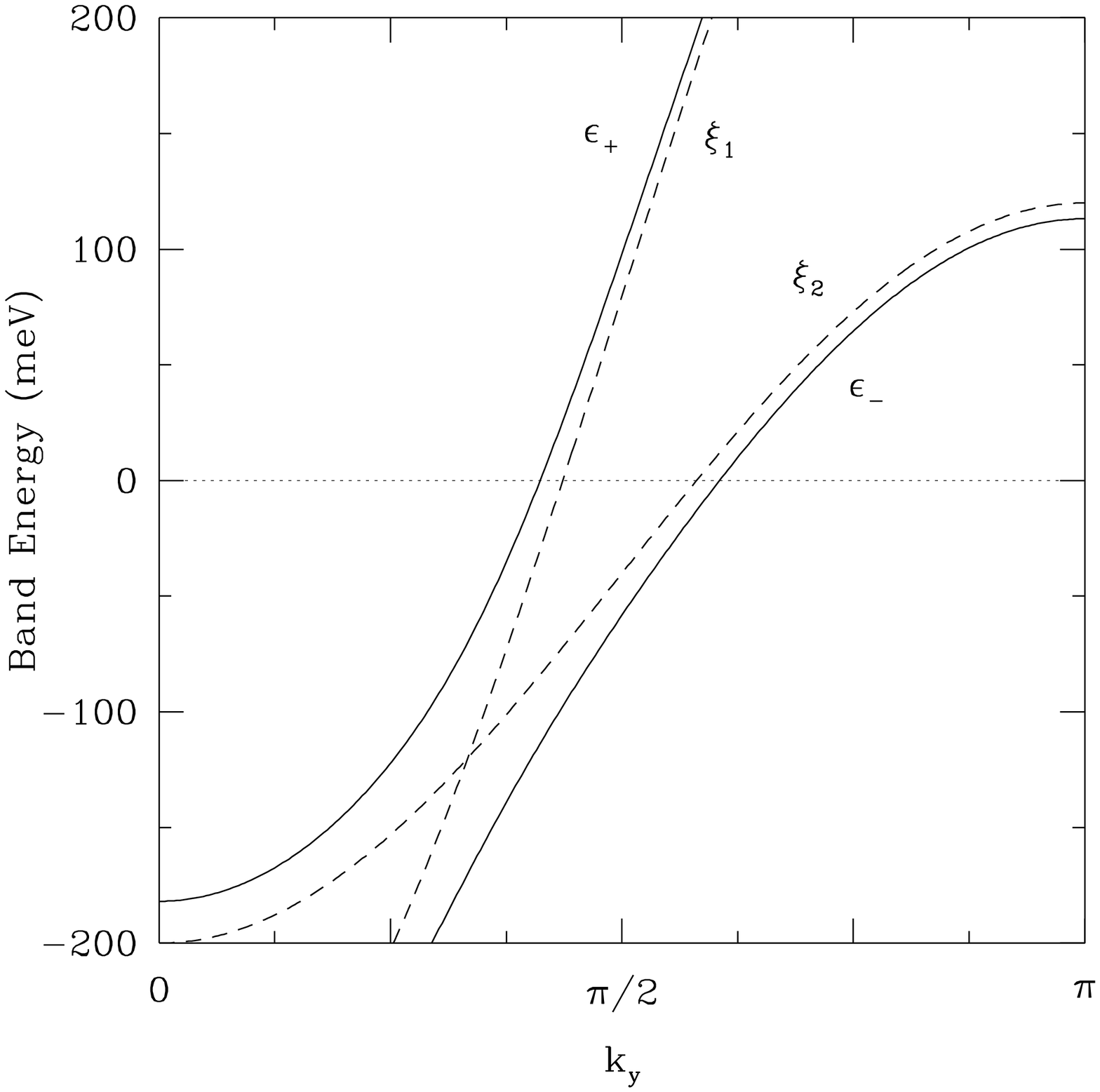}{0.4}{Energy
dispersions along the line $k_y = k_x$ for $k_z = 0$.  There is an
avoided crossing at $\xi_1 = \xi_2$.  The band parameters are typical
and are $t_1 = 100$ meV,
$t_2 = 80$ meV, $\mu_1 = -80$ meV, $\mu_2 = 40$ meV and $t_\perp
=25$ meV.}
\myfigure{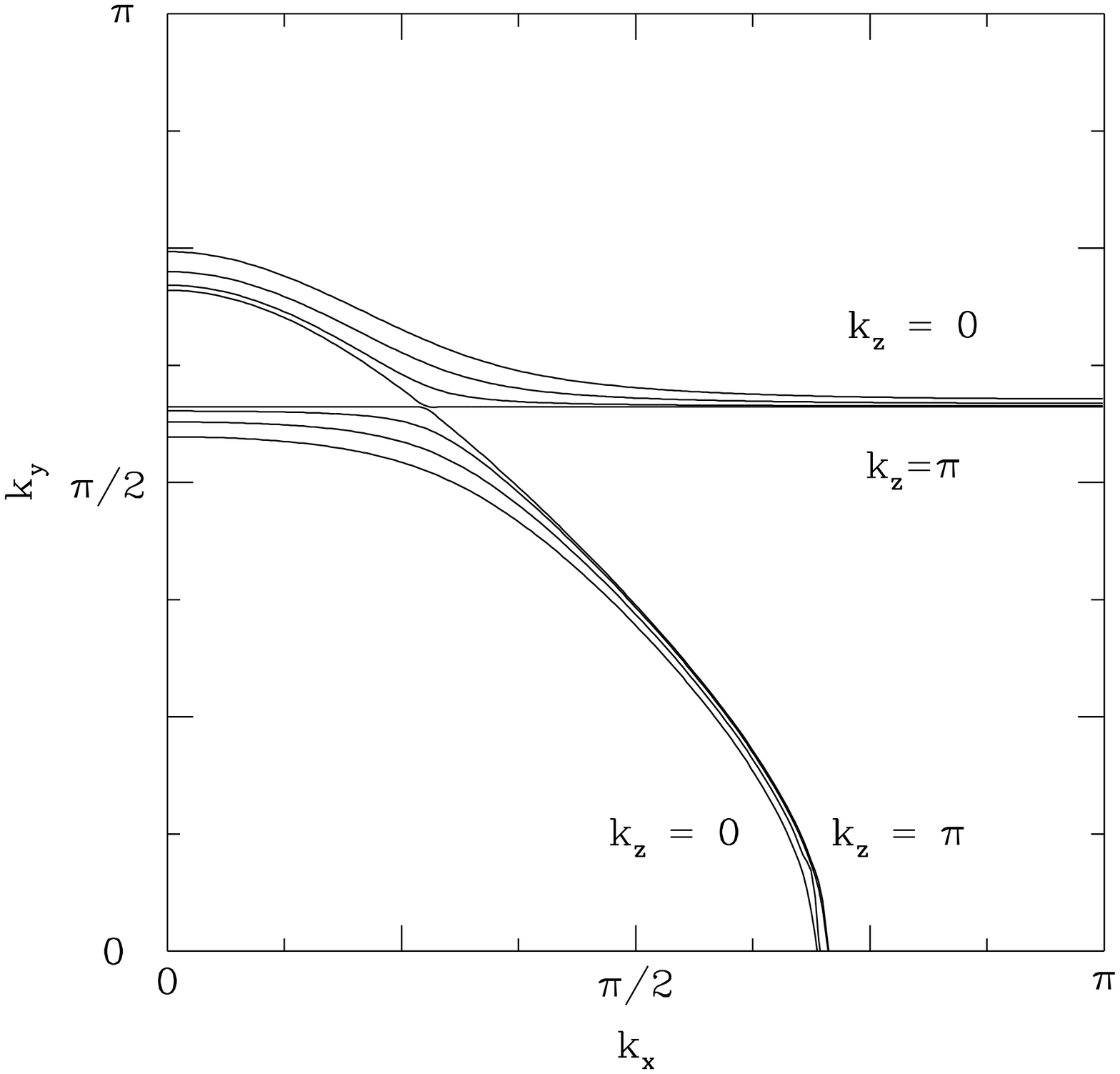}{0.4}{The Fermi surface is shown for a 
range of $k_z$ between $k_z = 0$ and $k_z = \pi$.  
When $k_z=\pi$, the chain--plane coupling vanishes and
the two pieces of Fermi surface are those of the isolated chains and
planes.  As the chain--plane coupling increases, the Fermi surfaces 
hybridize and are pushed apart.  The amount of hybridization at a
given ${\bf k}$ depends on the relative sizes of $t_\perp^2$ and 
$(\xi_1-\xi_2)^2$.
There is an avoided crossing of the Fermi surfaces when $\xi_1 = \xi_2
 = 0$. }
\myfigure{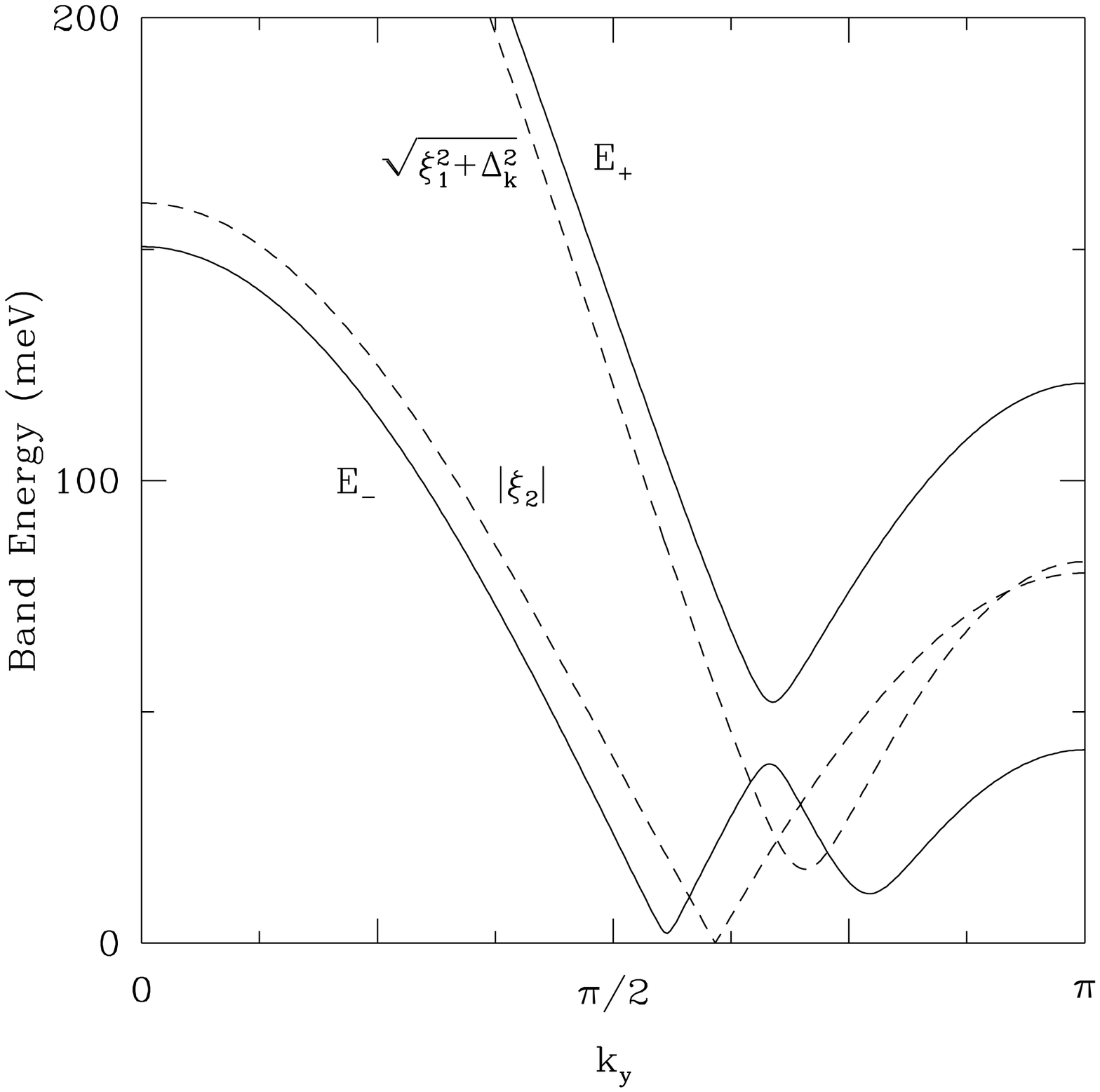}{0.4}{The quasiparticle
excitation spectra $E_\pm$ are plotted along the line $k_x = 0$ (solid
curves).  For these curves, $t_\perp =20$ meV. The excitation spectra
in the decoupled ($t_\perp = 0$) limit are plotted for comparison
(dashed curves).}
\myfigure{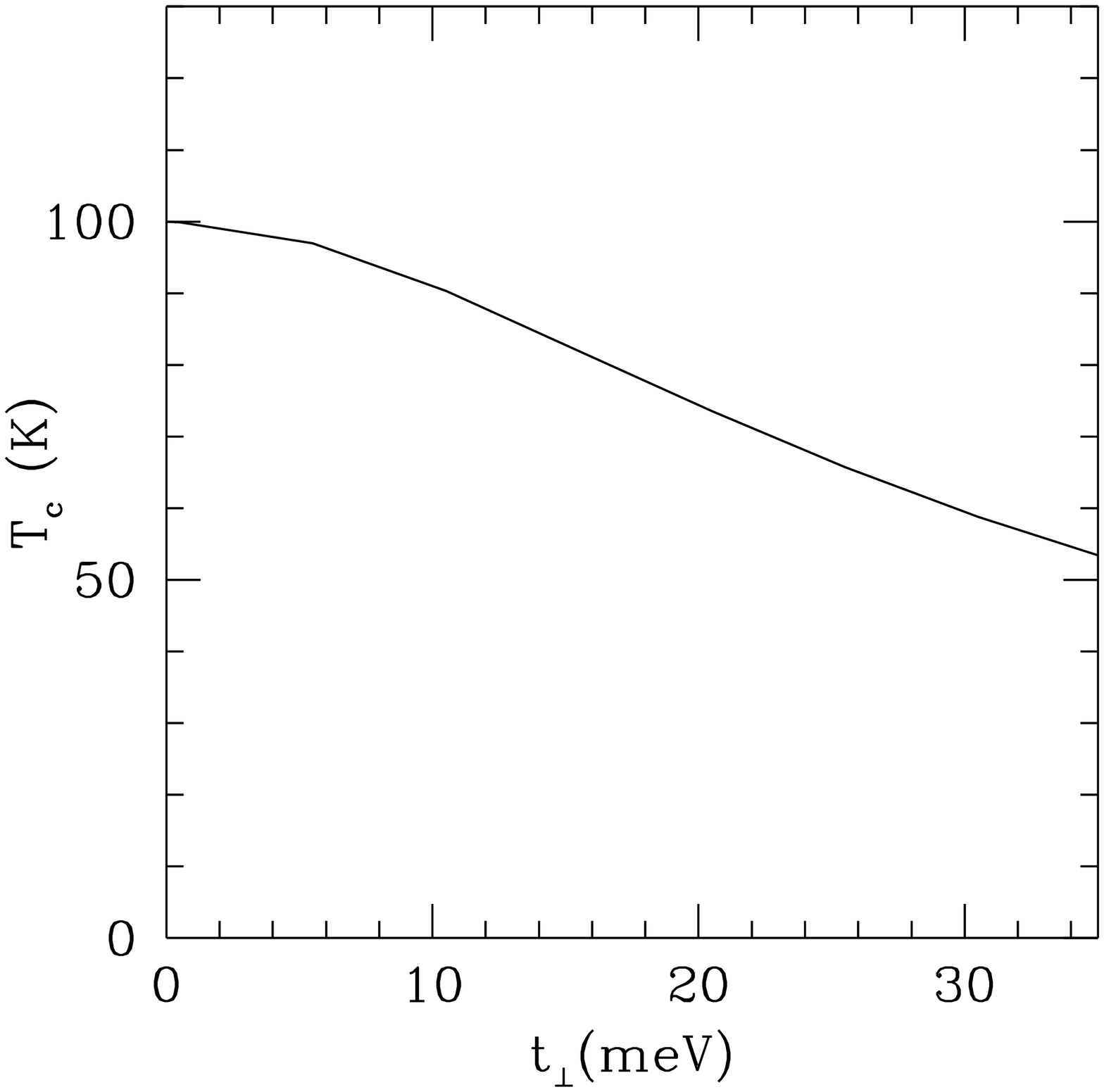}{0.4}{The critical
temperature in the intrinsically superconducting planes is lowered
by the proximity of the intrinsically normal chains. }
\myfigure{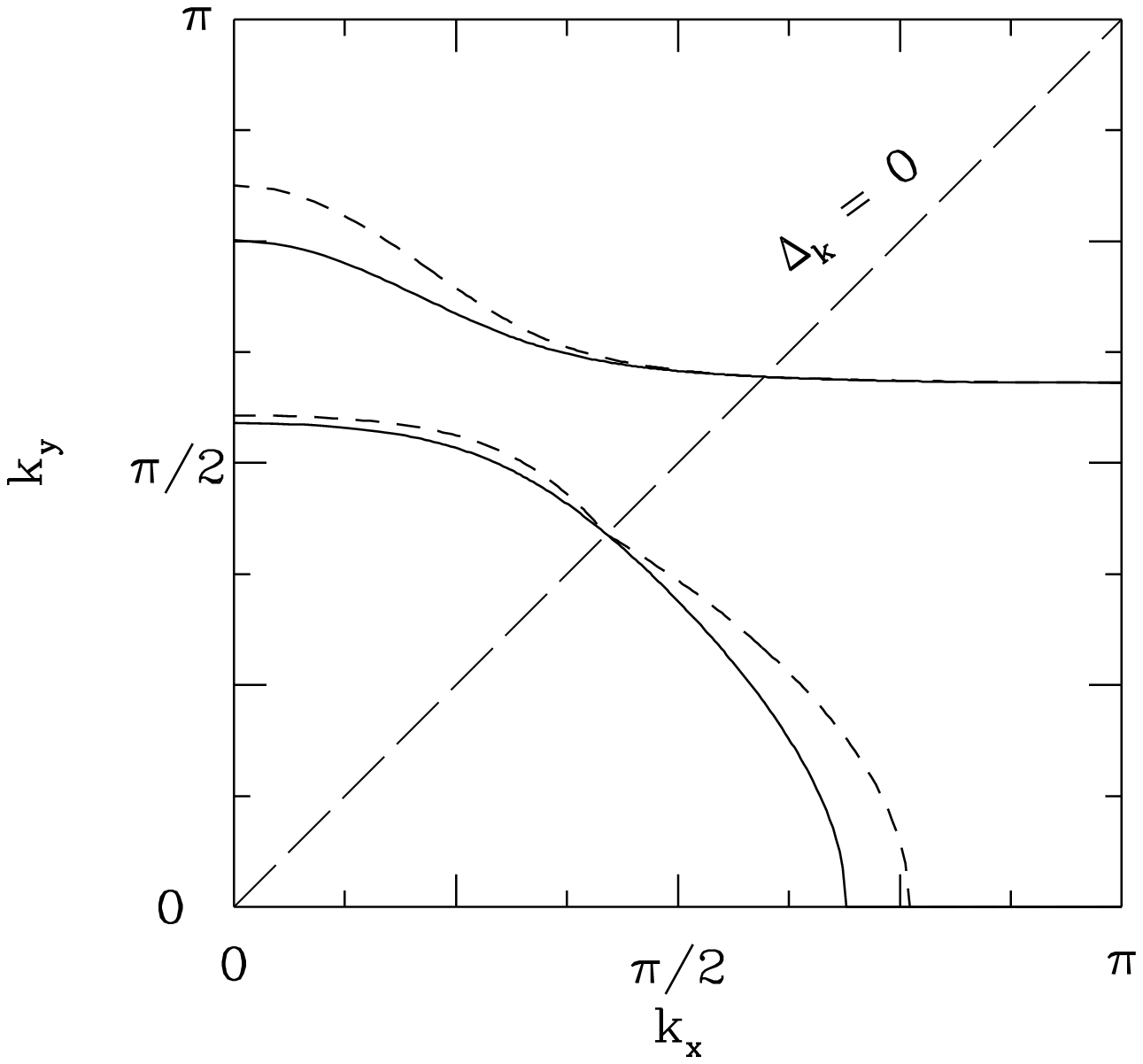}{0.4}{The value of the lower
energy band $E_-(k)$ (dashed line) is plotted along the Fermi surface
(solid line) of the chain-plane model for $k_z=0$.  The chain-plane
coupling strength is $t_\perp = 25$ meV. }
\myfigure{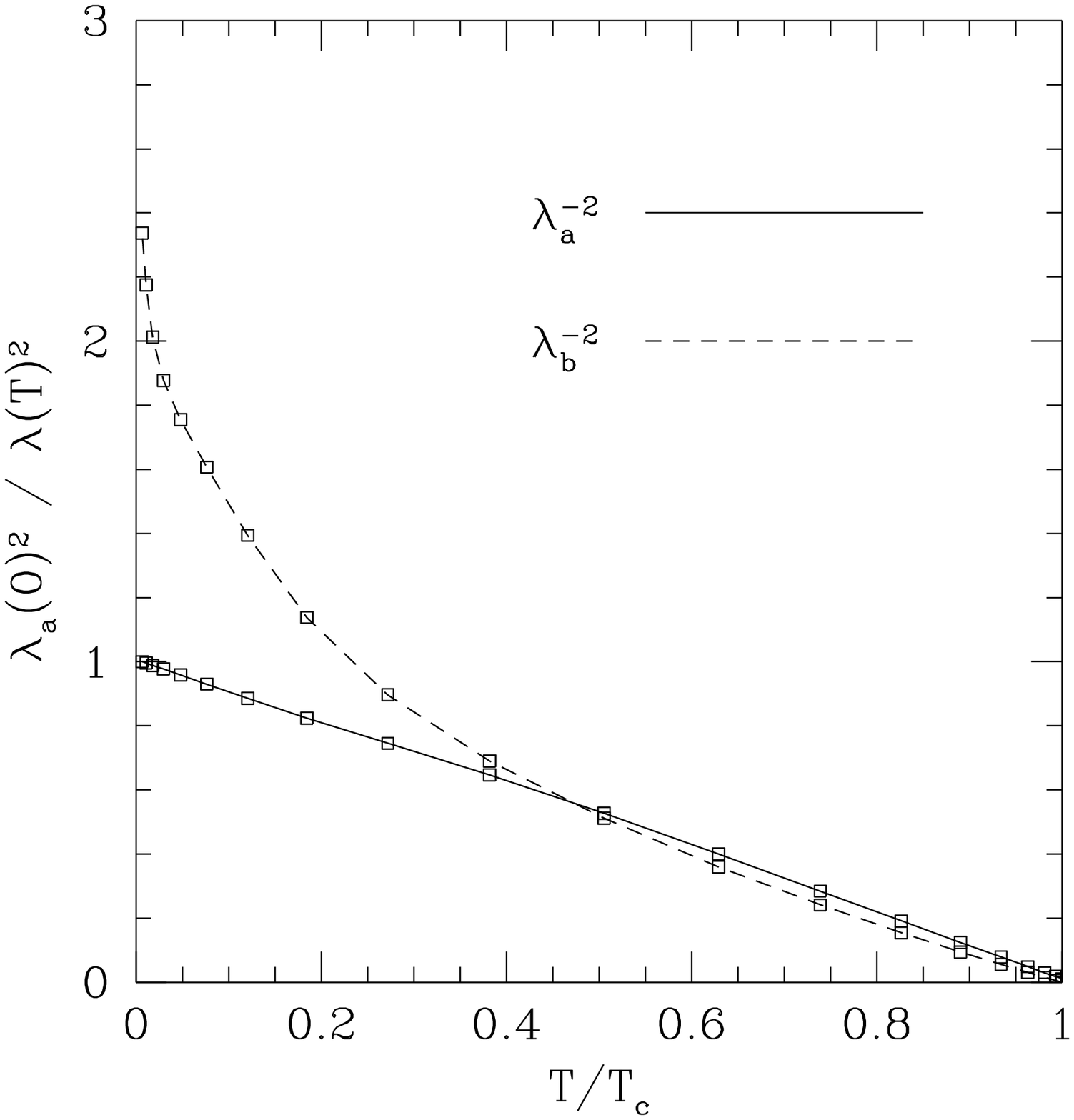}{0.4}{In--plane penetration depth for a $d$-wave
order parameter.  the penetration depth in the $a$ direction
(perpendicular to the chains) is nearly that of a pure $d$ wave superconductor
in the absence of chains.  The penetration depth in the $b$ direction
has a very different temperature dependence from that in the $a$ direction
because the size of the induced gap in the chains is much different from the
size of the gap in the planes.  The relative bandwidths of the
chains and planes were determined by setting $\lambda_a^2(0)\lambda_b^2(0)
\sim 2.5$, in accordance with experiment.}
\myfigure{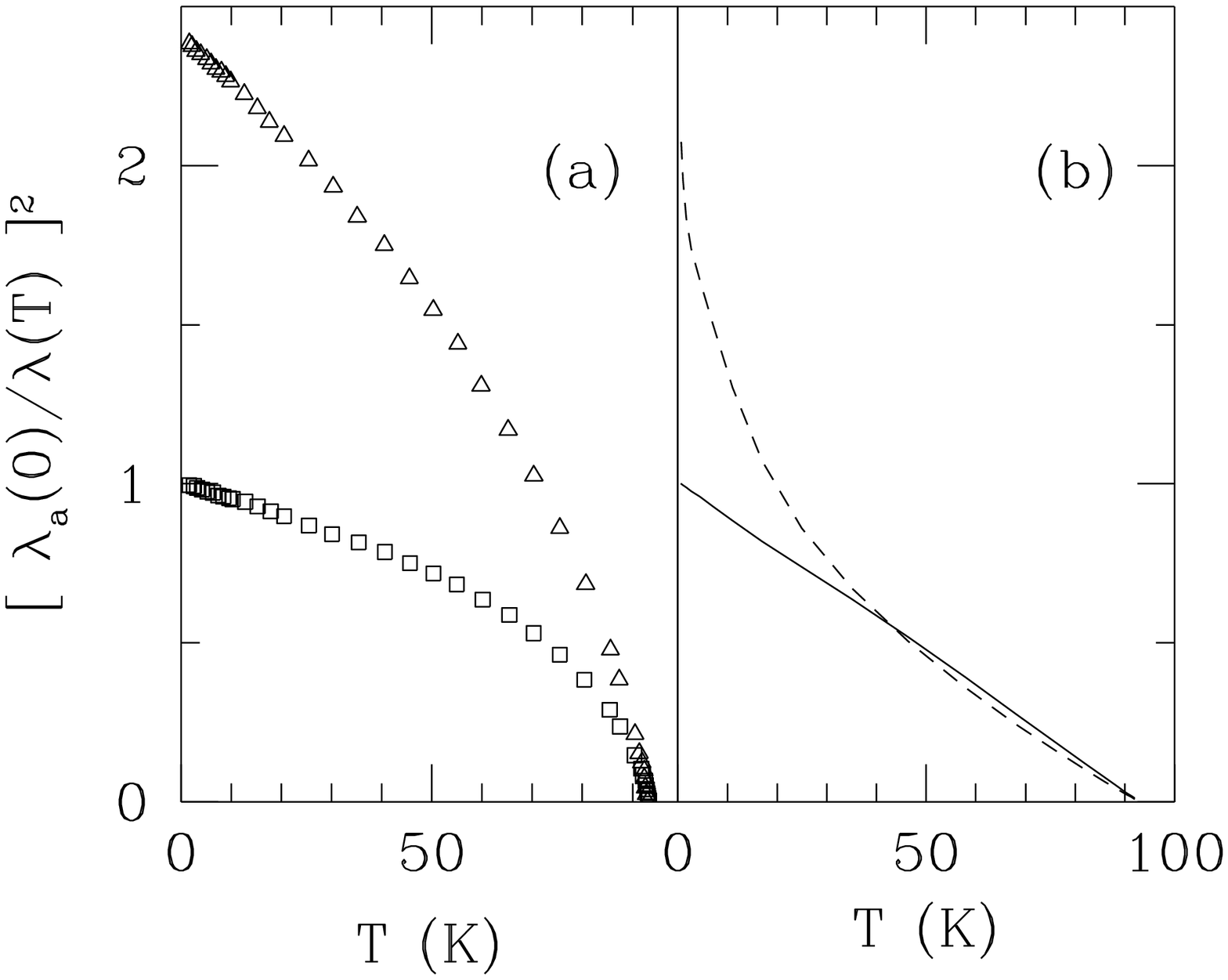}{0.55}{A comparison of penetration depths 
found by experiment and by the proximity model.
The experimental data \protect\cite{bonn2} on untwinned crystals
of YBCO is shown in (a).  Squares are $\lambda_a(0)^2/\lambda_a(T)^2$
and triangles are $\lambda_a(0)^2/\lambda_b(T)^2$.  In (b), the theoretical
values of the penetration depth are shown for the proximity model in
the $a$ (solid curve) and $b$ (dashed curve) directions.}

A first model, which can be used to get some insight into
the situation for optimally doped YBCO, is that of planes and
chains coupled through a transverse tunnelling matrix element $t_\perp$ with
the pairing interaction assumed to reside exclusively in the CO$_2$
plane.  The simplest electronic dispersion relations for such a
model \cite{atkinson1,atkinson2,atkinson3} are:
\begin{eqnarray}
\xi_1 &=& -2t_1 [\cos(k_x) + \cos(k_y)] - \mu_1\\
\xi_2 &=& -2t_2\cos(k_y) - \mu_2,
\end{eqnarray}
where $t_1$ is the first neighbour hopping in the CuO$_2$ plane and
$t_2$ in the CuO chains with $\mu$ the chemical potential.  Application
of an interplane matrix element $t_\perp$ will mix the
plane and chain band with resulting band energies having the form:
\begin{equation}  
\epsilon_\pm = \frac{\xi_1+\xi_2}{2} \pm \sqrt{\left(\frac{\xi_1-\xi_2}{2}
\right )^2 + t_\perp(\kk)^2},
\end{equation}
with
\begin{equation}
t_\perp(\kk) = 2 t_\perp \cos(k_z/2)
\end{equation}
in these relationships $k_x$, $k_y$ and $k_z$ are components of
momentum ranging from $(-\pi,\pi)$ in units of the inverse of the
lattice parameters.  Results are shown in Fig.\ 1.  As seen, the
effect of the plane-chain coupling is largest where $\xi_1(\kk) =
\xi_2(\kk)$ at which point there is an avoided crossing.  In Fig.~2,
we show how the Fermi surfaces are pushed apart in $k$-space and see
that the amount of distortion of the Fermi surfaces depends on their
proximity to the avoided crossing in the 2-D Brillouin zone.  The
various contours are for different values of $k_z$ with $k_z=\pi$
corresponding to no coupling ($t_\perp(k_z) = 0$) and $k_z$ = 0
corresponding to the largest effect.  The area between these two outer
contours corresponds to the Fermi surface dispersion in the
z-direction.  If it was a perfectly flat cylinder--like structure in
this direction, it would project into a single curve in the 2-D
Brillouin zone.

One result of the large critical temperature in the oxides is that they
have an extremely small
coherence length $\xi_0$.  The coherence length is the distance scale
over which the superconducting order parameter $\Delta$ may vary
spatially.  In the {\sc bcs} theory it is :
\[
  \xi_0 = \frac{\hbar v_f}{\pi \Delta},
\]
where $v_f$ is the electron Fermi velocity, and $\Delta$ is the
magnitude of the {\sc bcs} order parameter.  For a {\sc bcs}
superconductor with $T_c = 100 \mbox{ K}$, $\Delta = 15 \mbox{ meV}$.
The Fermi velocity can be estimated from the bandwidth of the
conduction band, and since the high $T_c$ materials are highly
anisotropic, there will be substantial differences between $v_f$ in
the various directions.  The coherence length will therefore be
anisotropic as well.  In the $a$ and $b$ directions (parallel to the
CuO$_2$ planes) the CuO$_2$ bandwidth is $t_{\|} \sim 1 \mbox{ eV}$.
The Fermi velocity can be estimated as:
\[
  v_{fa} \sim v_{fb} \sim \frac{a t_\|}{\hbar \pi}
\]
where $a$ is the lattice constant in the $a$-direction. 
For the BSCCO compounds, $a \sim 5$\AA.  The coherence length in the 
planes is therefore:
\[
  \xi_a \sim \xi_b \sim \frac{a t_{\|}}{\pi^2 \Delta} \sim 25{\rm \AA}.
\]
The bandwidth along the $c$ axis is considerably smaller than in the
CuO$_2$ planes.  In the BSCCO compounds, it is typically $t_\perp \sim
10 \mbox{ meV}$.  Since the unit cell size along the $c$-axis is $c
\sim 30$\AA, the coherence length in BSCCO is
\[
  \xi_c \sim \frac{c t_{\perp}}{\pi^2 \Delta} \sim 2{\rm \AA}.
\]
These values are typical for most of the high $T_c$ materials and
distinguish them from the conventional materials in one important way:
the fact that $\xi_c$ is substantially less than the length of the
unit cell along the $c$ axis allows the order parameter to vary
spatially {\em over the unit cell}.  In the conventional materials,
where the coherence lengths are $10^3$ or $10^4$\AA, the structure of
the unit cell is invisible to the gap.  In the high $T_c$ materials,
the value of the gap may depend on the layer type: chain or plane.

       Next we wish to include the superconductivity.  We will assume
that the pairing is operative only in the CuO$_2$ plane and that the
chains become superconducting only through the tunnelling matrix
element $t_\perp$.  In mean field theory, the Hamiltonian, $H$, is:
\begin{equation}
H - N\mu = \sum_\kk C^\dagger ({\kk})Q({\kk})
  { C}({\kk}) + \mbox{const.},
\end{equation}
with:
\begin{equation}
  \label{2.31}
  { C} ({\kk}) = \left [ \begin{array}{c}
  { c}_{1 {\kk} \uparrow} \\
  { c}^\dagger_{1 -{\kk} \downarrow} \\
  { c}_{2 {\kk} \uparrow} \\
  { c}^\dagger_{2 -{\kk} \downarrow} 
  \end{array} \right ],
\end{equation}
where $c^\dagger_{i\kk\sigma}$ 
are the creation operators for electrons in layer $i$ and
spin $\sigma$.  Here the matrix $Q(\kk)$ is $4\times4$ and equal to:
\begin{equation}
  \label{2.32}
  Q(\kk) =  \left [ \begin{array}{cccc}
  \xi_1({\kk}) & -\Delta_{ \kk} & t({ \kk}) & 0 \\
  -\Delta^\ast_{ \kk} & -\xi_1(-{ \kk}) & 0 & -t^\ast(-{ \kk}) \\
  t^\ast({ \kk}) & 0 &  \xi_2({ \kk}) & 0 \\
  0 &  -t(-{ \kk}) & 0 & -\xi_2(-{ \kk})
  \end{array} \right ] .
\end{equation}
The gap $\Delta_\kk$ is given by:
\begin{equation}
  \Delta_\kk  = \frac{1}{N} \sum_{\kk^\prime} V_{\kk\kk^\prime}
  \langle c_{1-\kk^\prime\downarrow} c_{1\kk^\prime\uparrow} \rangle,
\end{equation}
where $\Omega$ is the volume, $V_{\kk\kk^\prime}$ 
is the pairing potential in the
copper oxide plane denoted by the subscript $1$.  We will assume 
$V_{\kk\kk^\prime}$ 
to be separable and have $d$-wave symmetry.  That is:
\begin{equation}
V_{\kk\kk^\prime} = \eta_\kk V \eta_{\kk^\prime},
\end{equation}
with:
\begin{equation}
\eta_\kk = \cos(k_x) - \cos(k_y).
\end{equation}

Diagonalisation of the Hamiltonian (5) leads to four energy bands
$E_1 = E_+, E_2 = E_-, E_3 = -E_-, E_4 = -E_+$ with:
\begin{equation}
 E_{\pm} = \frac{\xi_1^2 + \xi_2^2 + \Delta_{\bf k}^2}{2} + t^2 
  \pm \sqrt{ \left [ \frac{\xi_1^2 + \xi_2^2 + \Delta_{\bf k}^2}{2} + t^2 
  \right ]^2
  - (t^2 - \xi_1 \xi_2)^2 - (\xi_2 \Delta_{\bf k})^2 }.
\end{equation}

In Fig.~3, we show the energy bands for $t_\perp = 0$ (dashed curves)
and $t_\perp = 20$ meV (solid curves).  For the uncoupled situation
shown for comparison (dashed curve), the chain band exhibits no
gap while the plane band shows the usual {\sc bcs} gap at its Fermi
surface.  Note that the curves are for $k_x = 0$ as a function of
$k_y$ and that the $d_{x^2-y^2}$ gap of the form (10) is nonzero along 
that line
in the two dimensional Brillouin zone.  When $t_\perp \neq 0$, the bands mix
as they would in the normal state with a band crossing.  In
addition, a gap is induced in the $E_-$ branch corresponding to
induced superconductivity in the chains coming from the proximity
matrix element $t_\perp$.  This gap is small compared with the value
of the original gap in the CuO$_2$ plane which is seen at somewhat
higher momentum in this figure.

       A first question that needs to be answered is how is $T_c$
changed when $t_\perp$ is switched on.  This is shown in Fig.~4 where
we have plotted the value of $T_c$ as a function of $t_\perp$ for a case
when the critical temperature $T_c = $100 K in the limit $t_\perp = 0$. 
We see that as $t_\perp$ (in meV) increases, $T_c$ is reduced by the
proximity of the chains but remains substantial even for a value of
$t_\perp = 20$ meV.

       In Fig.~5, we show the chain-plane Fermi surfaces (solid
curves) as contours in the first Brillouin zone for $k_z = 0$ and
$t_\perp = 25$ meV.  The value of $E_-(\kk)$ (dashed line) is also plotted
along the Fermi curves.  It is seen that in the region of the chain
Fermi surface which is not close to the plane Fermi surface, i.e.
the region around $(\pi,\pi/2)$ in the figure, the induced gap is small.  In
fact, it can be shown that in this region:
\begin{equation}
[\xi_1(\kk) - \xi_2(\kk)]^2 \gg t_\perp(\kk)^2,
\end{equation}
and that:
\begin{eqnarray}
\label{13}
E_-(\kk) &\sim& |\Delta_\kk| \frac{t_\perp^2}{(\xi_1-\xi_2)^2}\nonumber \\
& \sim & 0.1 \mbox{ meV},
\end{eqnarray}
for the parameters used in the calculations.  This small gap is an
important generic feature of a pure proximity model with a plane-%
chain Fermi surface (2 sheets).

       In a coupled two band model, the expression for the
electromagnetic response tensor $K_{\mu\nu}$ is complicated and has the
form \cite{atkinson2}:
\begin{equation}
K_{\mu\nu} = \frac{1}{c\Omega} \left \{ \left. G^j_{R\mu\nu}(0,0;0) 
\right|_{\Delta=0} - G^j_{R\mu\nu}(0,0;0) \right \},
\end{equation}
where:
\begin{eqnarray}
\label{15}
G^j_{R\mu\nu}(0,0;0) &=& -\frac{e^2}{\Omega \hbar^2 c} 
  \sum_{i,j = 1}^4 \sum_{\bf k} [\hat{\gamma}_\mu(\kk,\kk)]_{ij}
  [\hat{\gamma}_\mu(\kk,\kk)]_{ji} \nonumber \\
  &\times & \left [ \delta_{i,j} \frac{\partial f(E_i)}{\partial E_i} + 
  [1-\delta_{i,j}]\frac{f(E_i) - f(E_j)}
  {E_i - E_j} \right], 
\end{eqnarray}
where e is the charge on the electrons, $c$ is the speed of light,
$\Omega$ is the volume, $\hbar$ Planck's constant, and 
$[\hat{\gamma}_\mu(\kk,\kk)]_{ij}$ is the appropriate
electromagnetic vertex which, in the familiar one band model,
would be the Fermi velocity.  Here we have 
$[\hat{\gamma}_\mu(\kk,{\bf k+q})]_{ij} = U^\dagger(\kk) 
[{\gamma}_\mu(\kk,\kk+{\bf q})]_{ij}U(\kk+{\bf q})$
with $U(\kk)$ the $4\times 4$
unitary matrix that diagonalizes the Hamiltonian (\ref{2.32}).  The 
vertex $[{\gamma}_\mu(\kk,\kk)]_{ij}$ is
related to the dispersion curves in the bands:
\begin{equation}
\label{16}
[{\gamma}_\mu(\kk,\kk)]_{ij} = (-1)^{i-1} \nabla_\kk Q^0(\kk)_{ij}
\end{equation}
where $Q^0$ is the Hamiltonian matrix (Eq.\ (\ref{2.32})) in the 
normal state.
In the limit of a single band (\ref{15}) properly reduces to the familiar
result:
\begin{equation}
\label{17}
K_{\mu\nu} = -\frac{e^2}{c\Omega} \sum_\kk v_\mu(\kk) v_\nu(\kk)
\left [ \frac{\partial f(E_\kk)}{\partial E_\kk} -
\frac{\partial f(\epsilon_\kk)}{\partial \epsilon_\kk} \right ]
\end{equation}
where ${\bf v}(\kk)$ is the Fermi velocity and
$E_\kk = [\epsilon_\kk^2 + \Delta_\kk^2]^{1/2}$.  The second term in equation
(\ref{17}) is evaluated in the normal state and is related to the value of
the zero temperature penetration depth.

       Results for $\lambda_a^{-2}$ and $\lambda_b^{-2}$ coming from
the chains and planes in our model are shown in Fig.~6.  In our
notation, the chains are along the b-direction.  For the current along
a-direction, the penetration depth (solid curves) shows linear
dependence at low temperature as expected for d-wave.  This is to be
contrasted with the exponentially activated behaviour found for the
s-wave case.  For the currents in the b-direction (dashed curve), the
chains contribute significantly to the superfluid density and the
shape of the curve is very different.  It shows a strong upward
bending at low T which reflects the low energy scale noted in formula
(\ref{13}) for the value of the gap which comes from regions of the
chain Fermi surface well away from the band crossing point.  This
feature is generic to proximity models with pairing confined to the
CuO$_2$ plane.  The data of Bonn and Hardy\cite{bonn,hardy2,bonn2} in
pure single untwinned crystals of YBCO are shown in Fig.~7 where they
are compared with our results.  It is clear that no low energy scale
is observed in the data along the b-direction.  In fact, on normalized
plots, the observed temperature variation is very similar between a-
and b-directions with the slope in the b-direction slightly steeper.
This indicates clearly that the gap in the chain is large and that no
small energy scale exists.  To understand better how the low
temperature slope is related to the d-wave gap, we show, in Fig.~8,
results for the penetration depth in a one band model with gap
$\frac{1}{2}\Delta_\circ[\cos(k_x) - \cos(k_y)]$ for several values of the
ratio $2\Delta_\circ/k_BT_c$.  In obtaining these results, a {\sc bcs}
temperature variation was taken for the temperature dependence of the
gap.  It is clear that small gap values give a curve with concave
upward curvature while for large values the curve is concave downward.
For {\sc bcs} $2\Delta_0/k_BT_c = 4.4$ and the curve is nearly a
straight line with $45^\circ$ slope.  If for the same zero temperature
gap value the critical temperature is decreased, the curve will
clearly have a smaller slope.  This is because in that case the system
at low temperature is expecting that it should have a straight line
behaviour with an intercept at much higher temperature.

       An analytic result that can be proved for a simple d-wave
model with circular Fermi surface in the 2-dimensional Brillouin
zone and gap variation of the form
\begin{equation}
\mit\Delta = \mit\Delta_\circ \cos(2\phi)
\end{equation}
where $\phi$ is an angle along the Fermi surface is that
\begin{equation}
\lambda^{-2}(T) = \frac{4\pi n e^2}{mc^2}
\left[ 1 - \frac{2\ln (2)}{\mit\Delta_\circ}k_{\rm B}T\right],
\end{equation}
for $T \rightarrow 0$.
This shows that the slope as a function of reduced temperature is
inversely proportional to the ratio $2\mit\Delta_\circ/k_{\rm B}T$.
As this ratio increases, the slope becomes less steep.  Here $n$ is
electron density and $m$ is electron mass.

\section*{Interband Pairing}

\myfigure{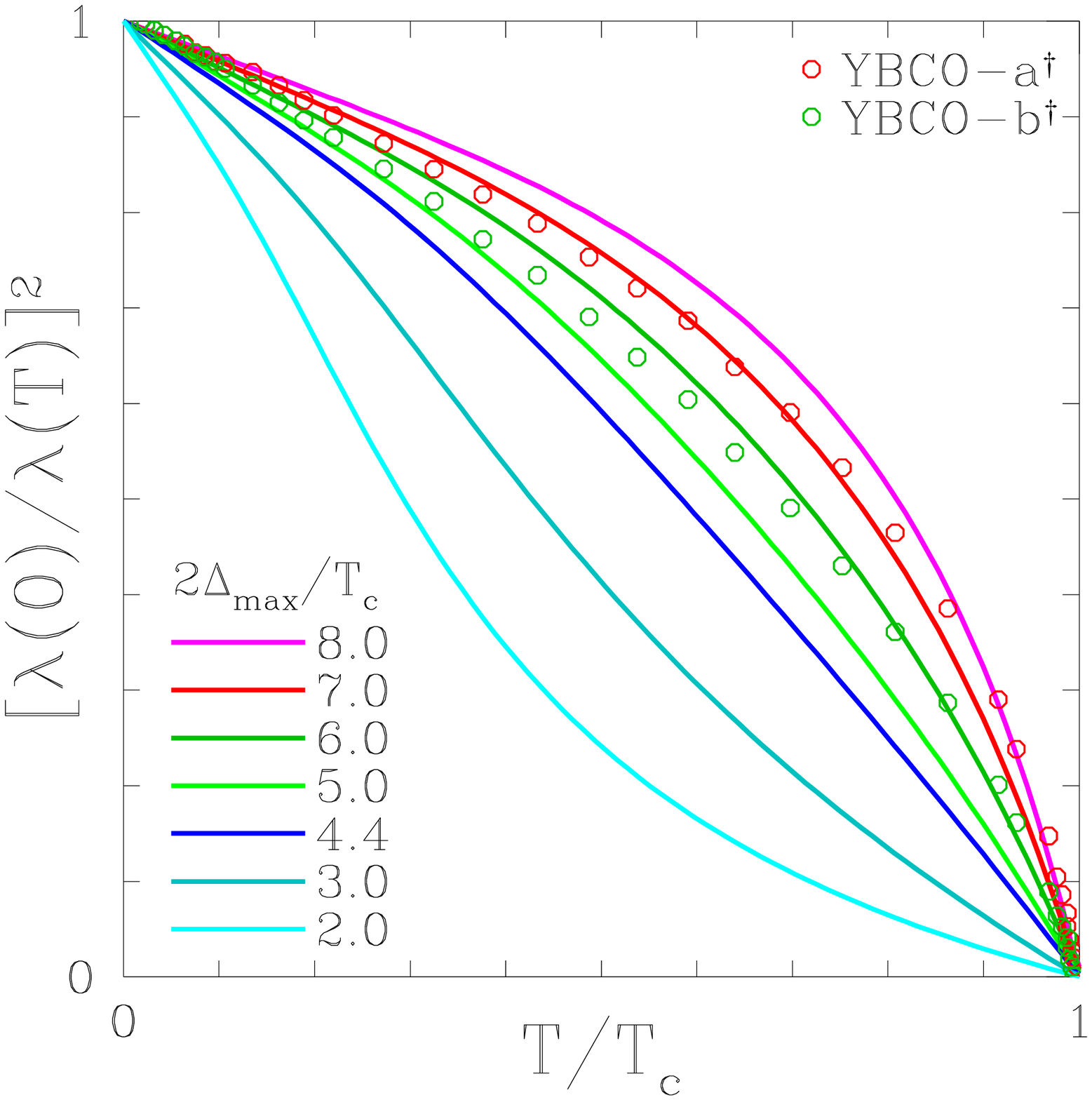}{0.4}
{The penetration depth, $[\lambda(0)/\lambda(T)]^2$, for different
values of $2\mit\Delta_{\rm max}/T_c$. From top to bottom the values
are: 8, 7, 6, 5, 4.4, 3 and 2. The data points are from microwave
experiments.\protect\cite{bonn2}}

\myfigure{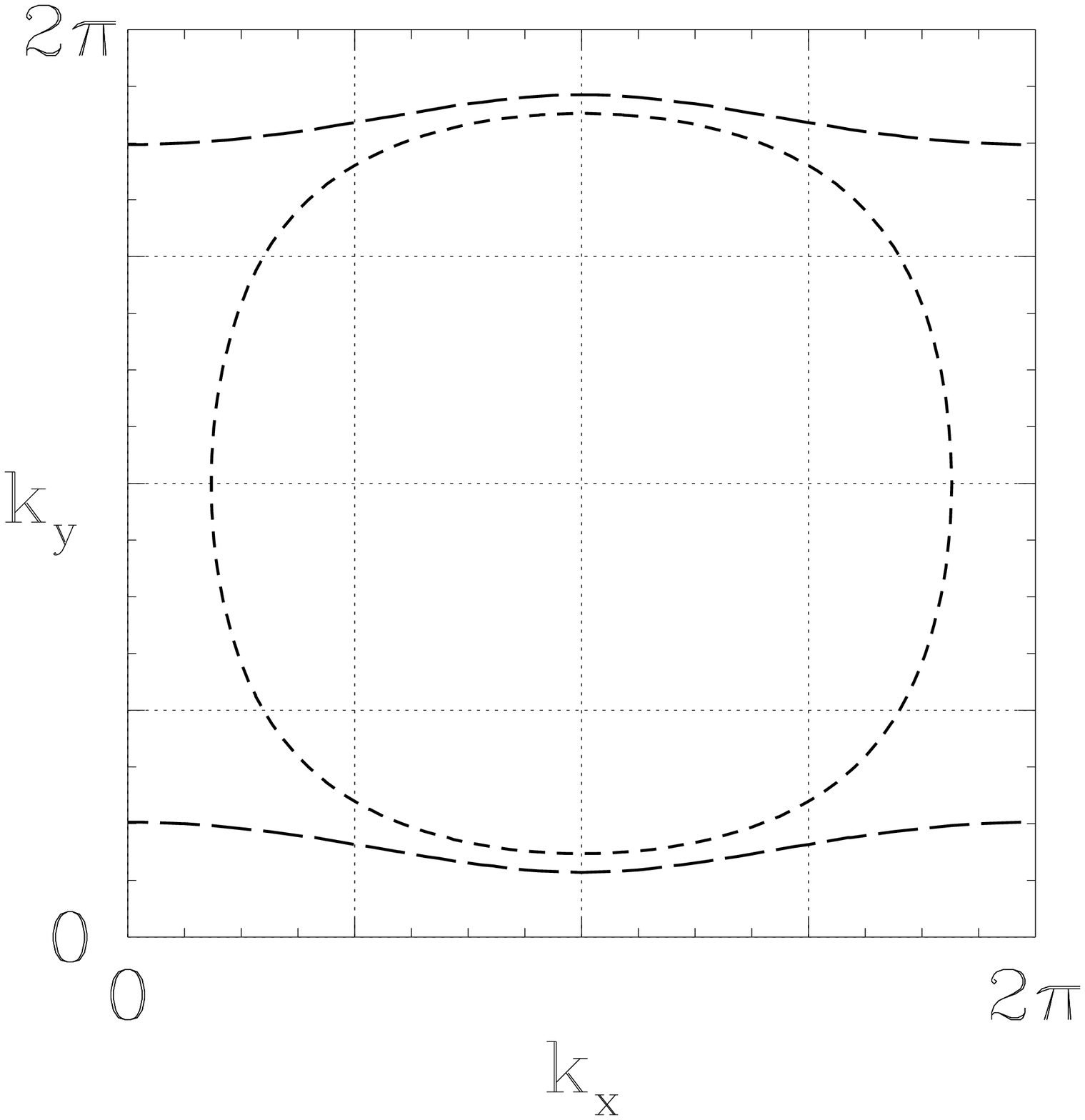}{0.4}
{Model of YBCO Fermi surfaces for chains (long dashed curve) and
planes (closed short dashed curve) in the first Brillouin zone. The
$(\pi,\pi)$ point is at the center of the figure.  For the chains the
parameters $\{t_\alpha,\epsilon_\alpha,B_\alpha,\mu_\alpha\}$ in
Eq.~\protect\ref{disp.eq} are $\{-50,-0.9,0,1.2\}$ and for the planes
they are $\{100,0,0.45,0.51\}$.}

%\begin{wrapfigure}{t}{0.75\columnwidth}
\begin{figure}[t]
\begin{center}
\begin{boxit}
{\setlength\tabcolsep{0pt}
\begin{tabular}{c c c c c}
	\makebox[0.3\columnwidth][l]{\large (a)} & &
	\makebox[0.3\columnwidth][l]{\large (c)} & &
	\makebox[0.3\columnwidth][l]{\large (e)}\\
                \epsfxsize=0.3\columnwidth
                \epsfbox{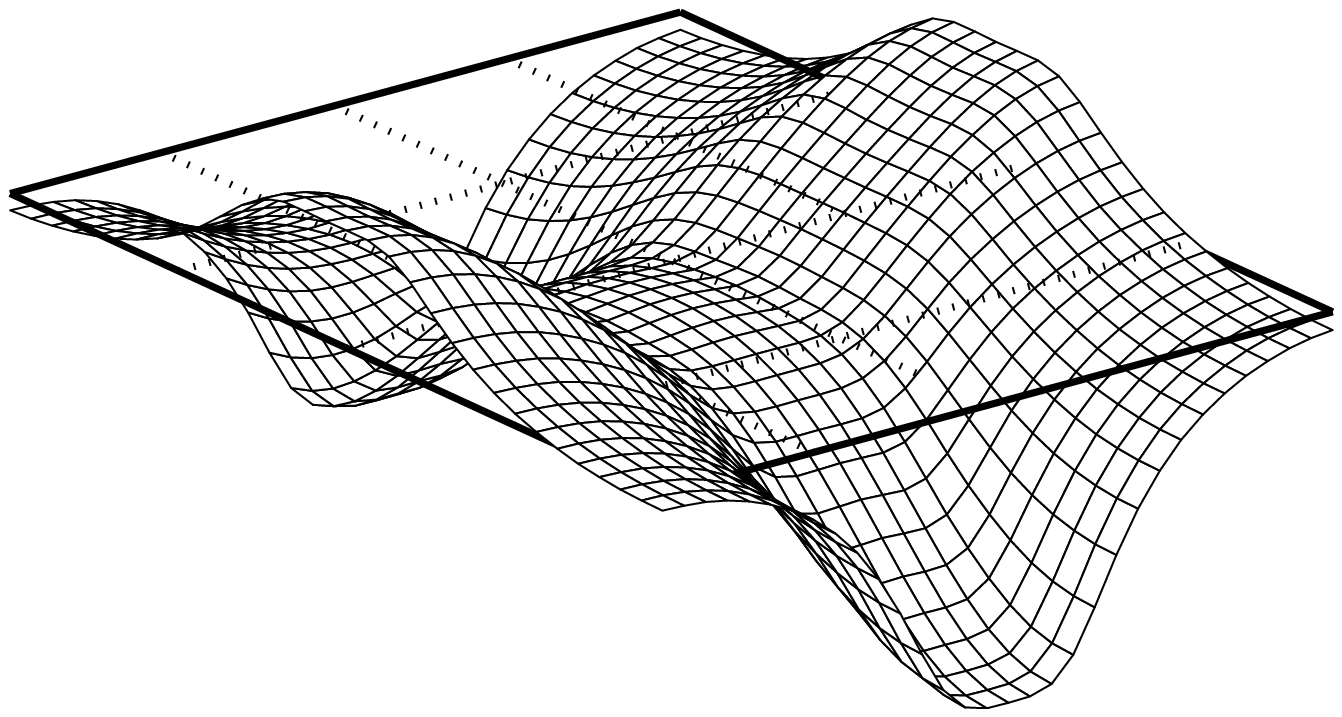} &
\parbox{10pt}{=\rule[-1.5cm]{0cm}{1.5cm}}&
                \epsfxsize=0.3\columnwidth
                \epsfbox{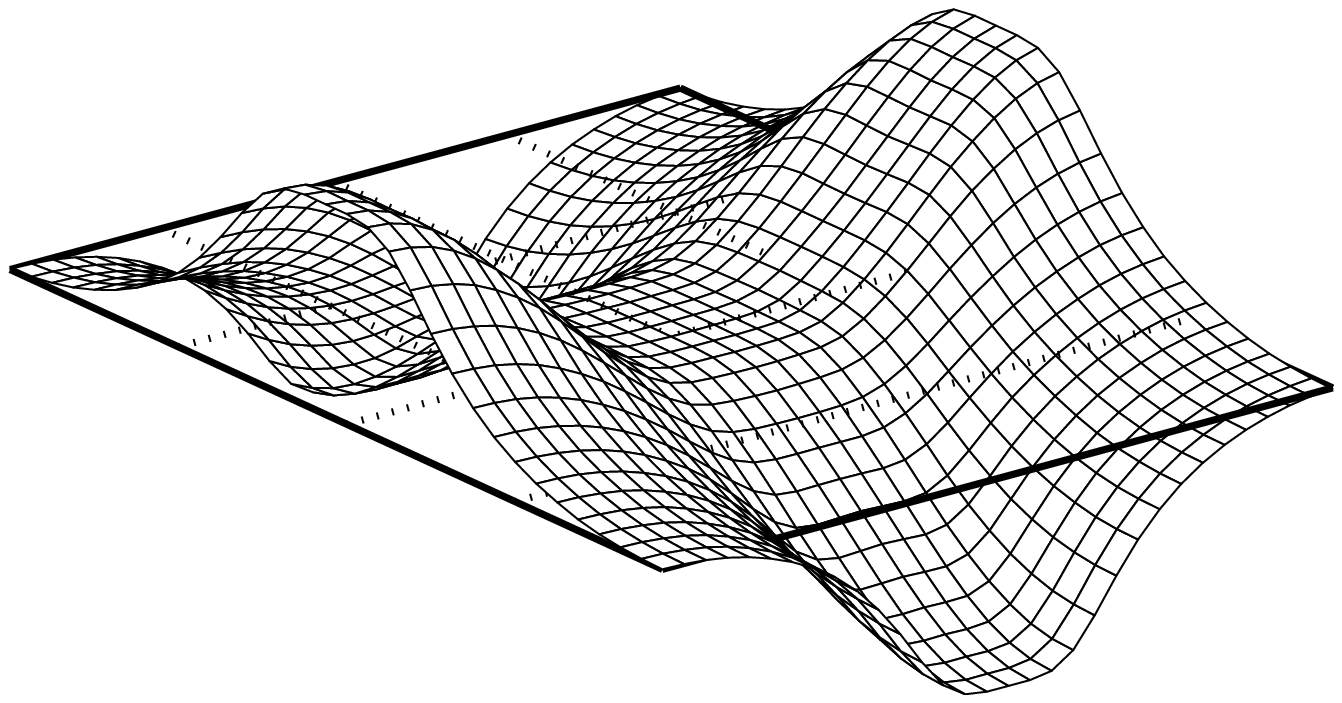} &
\parbox{10pt}{+\rule[-1.5cm]{0cm}{1.5cm}}&
                \epsfxsize=0.3\columnwidth
                \epsfbox{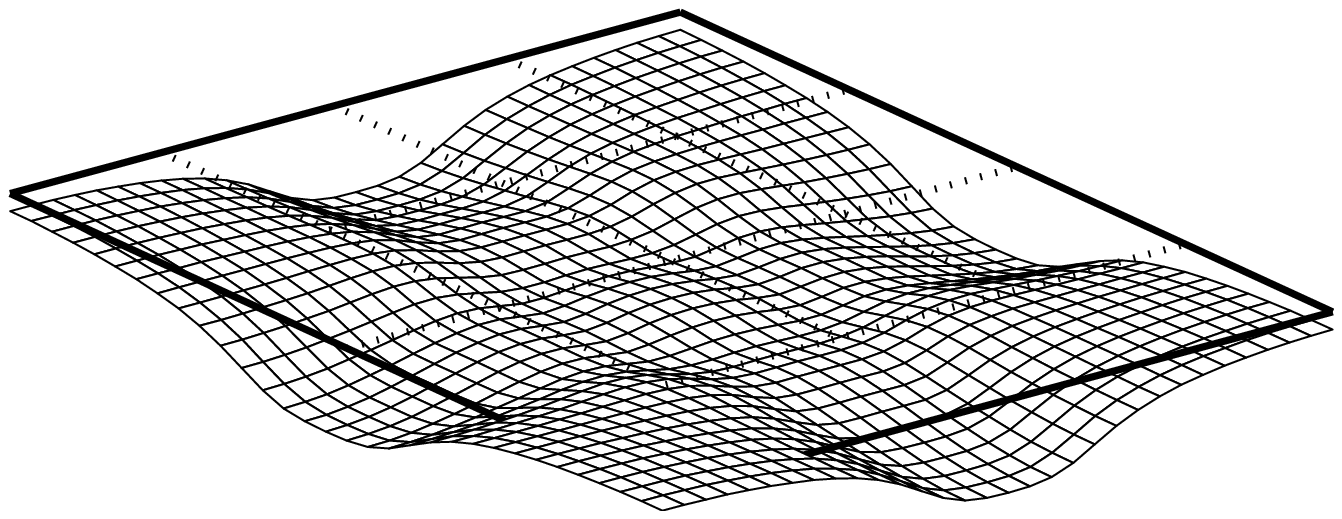} \\

	\makebox[0.3\columnwidth][l]{\large (b)} & &
	\makebox[0.3\columnwidth][l]{\large (d)} & &
	\makebox[0.3\columnwidth][l]{\large (f)} \\
                \epsfxsize=0.3\columnwidth
                \epsfbox{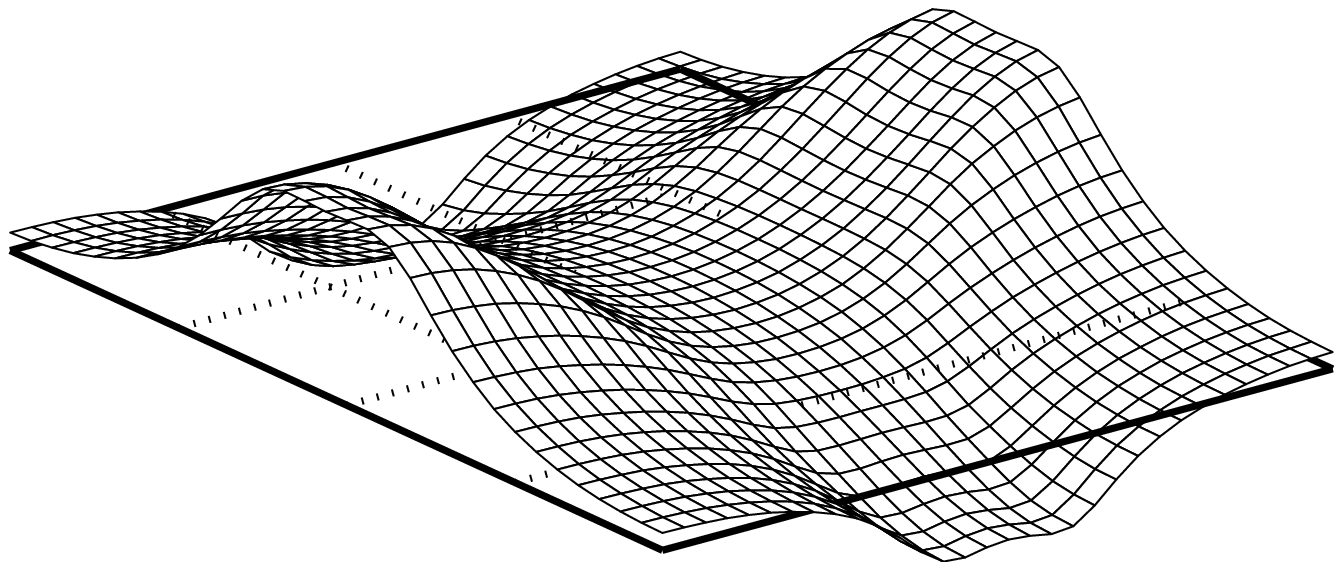} &
\parbox{10pt}{=\rule[-1.5cm]{0cm}{1.5cm}}&
                \epsfxsize=0.3\columnwidth
                \epsfbox{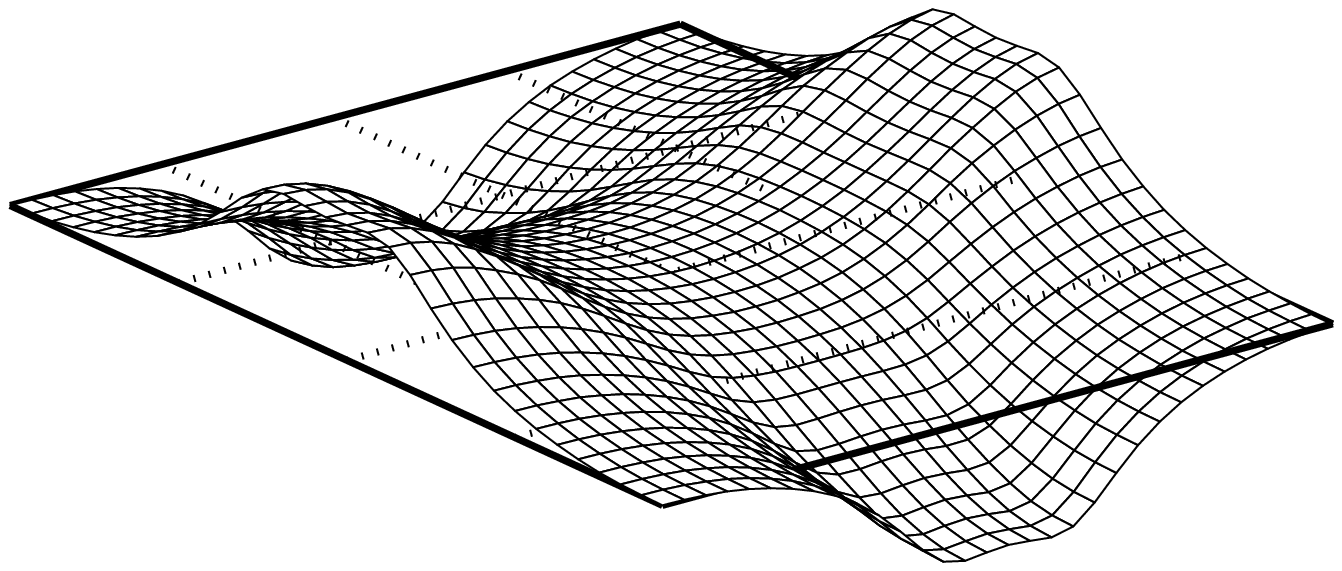} &
\parbox{10pt}{+\rule[-1.5cm]{0cm}{1.5cm}}&
                \epsfxsize=0.3\columnwidth
                \epsfbox{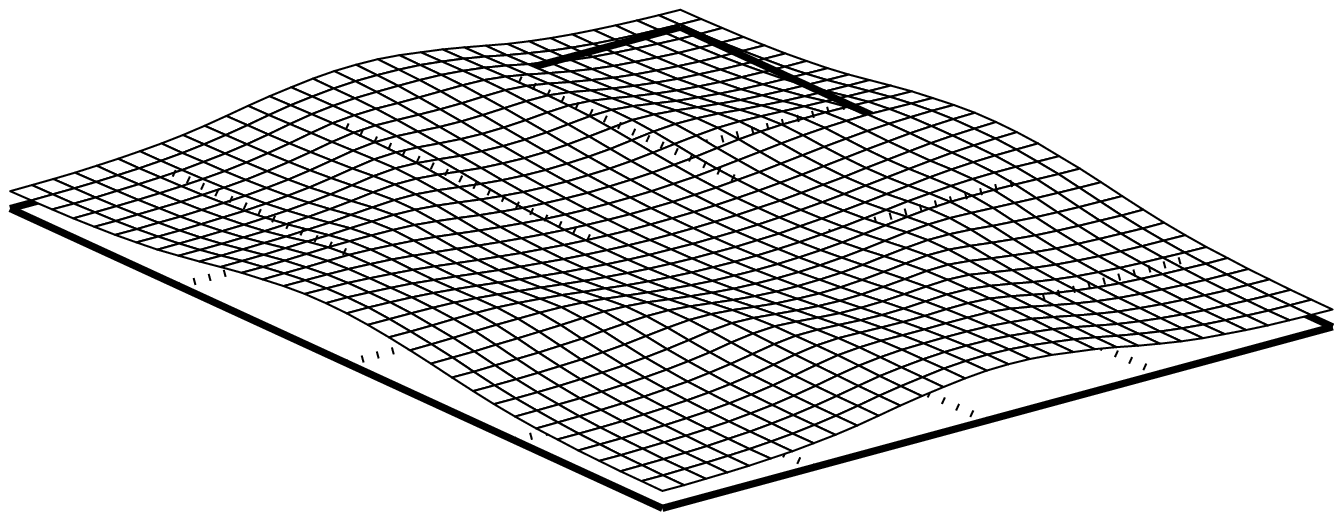}
\end{tabular}
}
\caption{\small
The order parameter in the first Brilliouin zone for (a) the plane
layers and (b) the chain layers. Beside are the projections of the $d$
components, (c) and (d), and the $s$ components, (e) and (f). The
vertical scale in all frames is the same. Note that the relative phase
of the $d$-components in the two layers are the same while that of the
$s$-components are opposite; this is caused by the interlayer
interaction, $V_{{\bf k},{\bf q},12}$, being negative (ie,
repulsive).}
\label{fig10.ps}
\end{boxit}
\end{center}
\end{figure}
%\end{wrapfigure}

%\begin{wrapfigure}{t}{0.99\columnwidth}
\begin{figure}[t]
\begin{boxit}
	\begin{center}
	\begin{tabular}{c c c} %\hline
			\makebox[0.25\columnwidth][l]{\large (a)} &
			\makebox[0.25\columnwidth][l]{\large (c)} &
			\makebox[0.25\columnwidth][l]{\large (e)} \\
			\epsfxsize=0.25\columnwidth
			\epsfbox{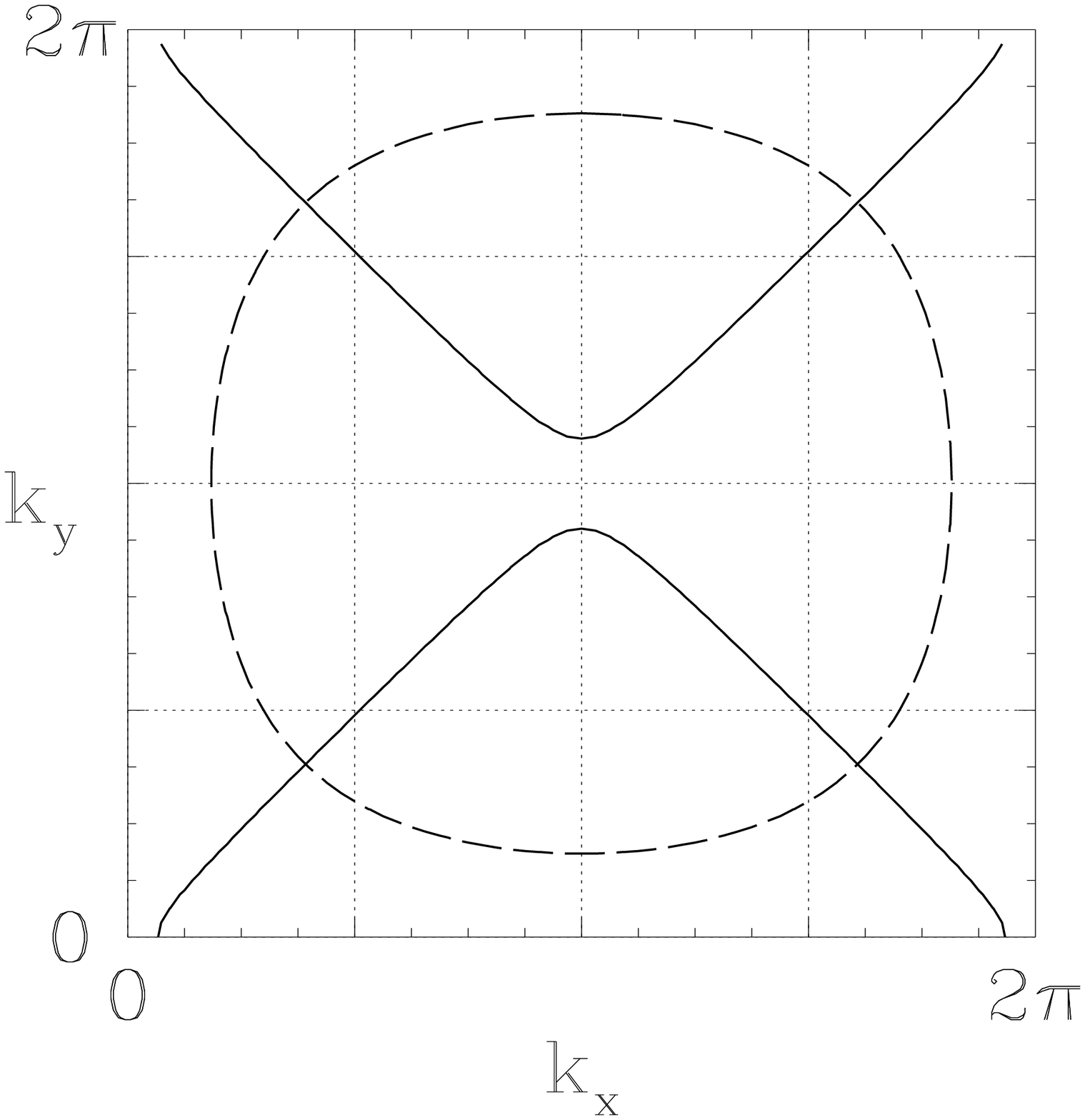} &
			\epsfxsize=0.25\columnwidth
			\epsfbox{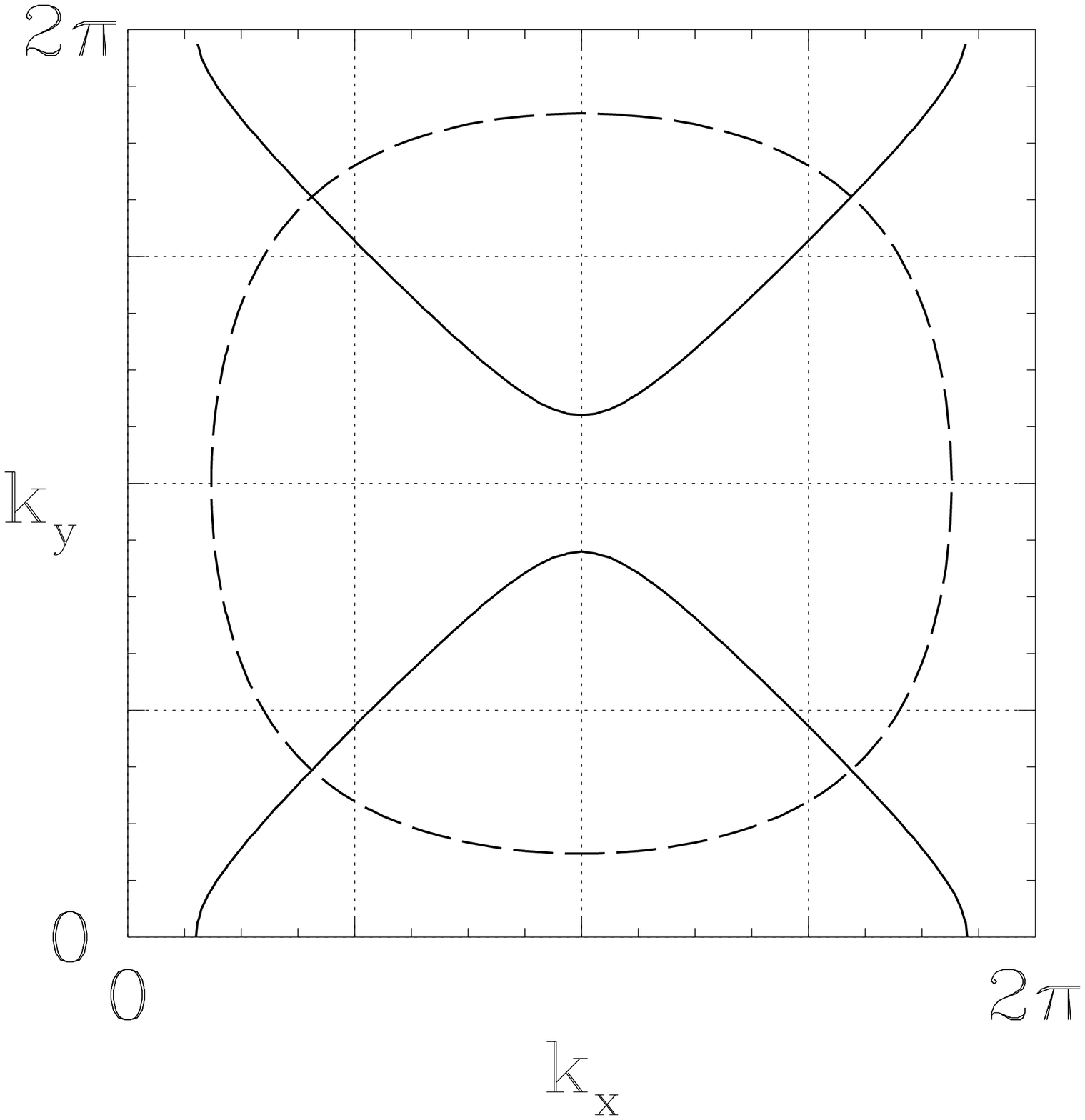} &
			\epsfxsize=0.25\columnwidth
			\epsfbox{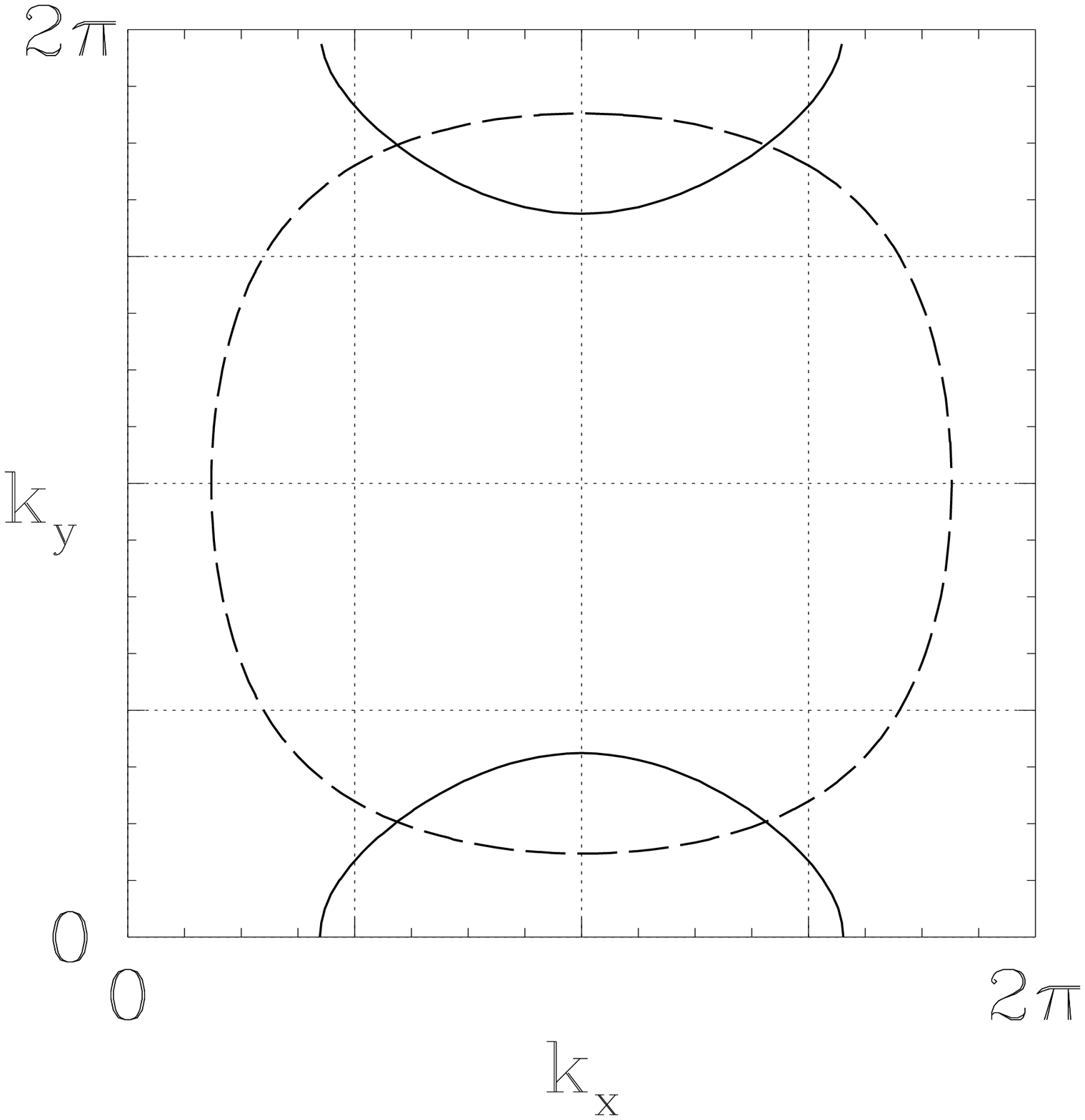} \\ %\hline
			\makebox[0.25\columnwidth][l]{\large (b)} &
			\makebox[0.25\columnwidth][l]{\large (d)} &
			\makebox[0.25\columnwidth][l]{\large (f)} \\
			\epsfxsize=0.25\columnwidth
			\epsfbox{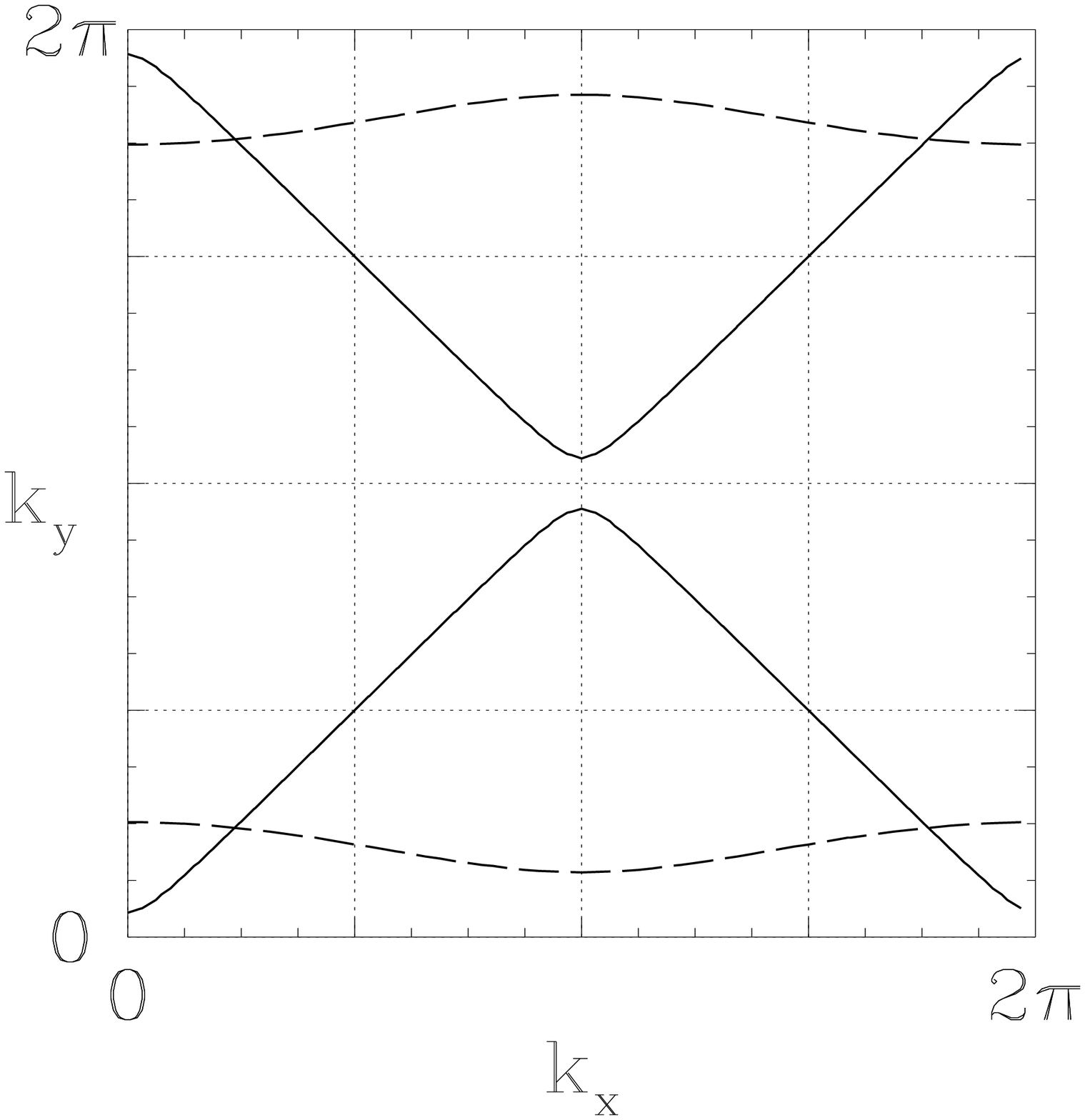} &
			\epsfxsize=0.25\columnwidth
			\epsfbox{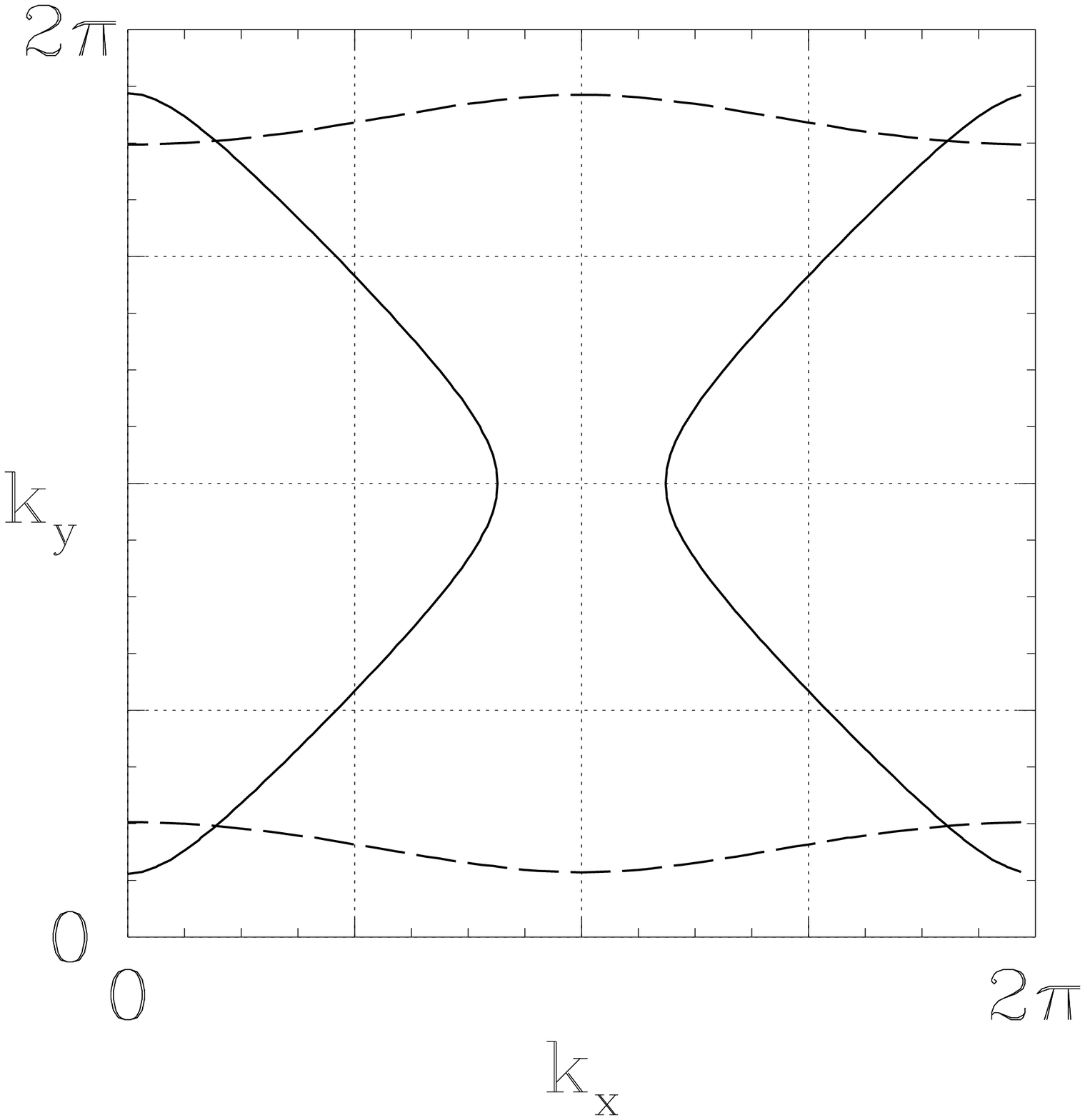} &
			\epsfxsize=0.25\columnwidth
			\epsfbox{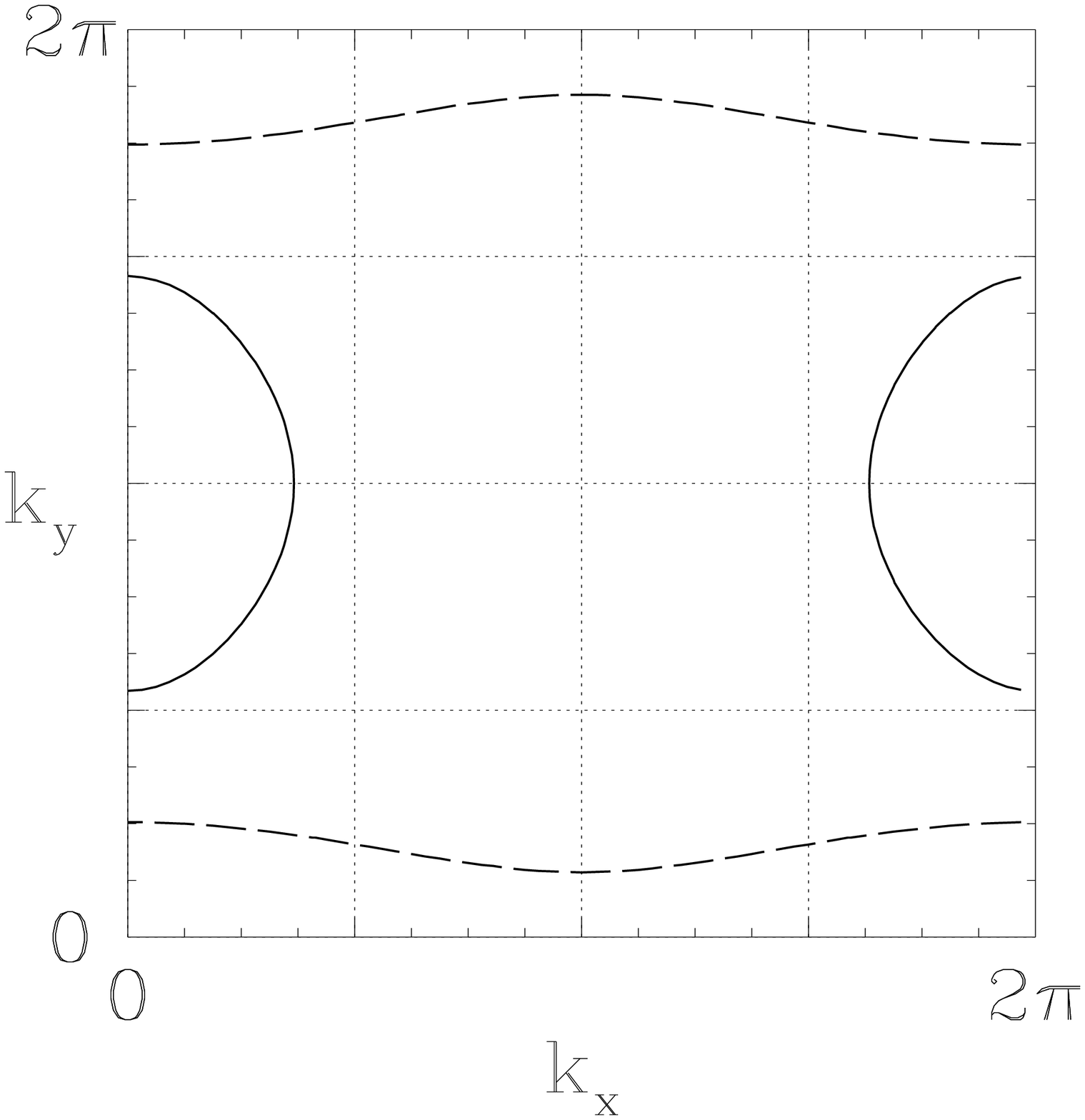}
		\end{tabular}
	\end{center}
\caption{\small
The Fermi surface (dashed curves) and gap nodes (solid curves) for a
CuO$_2$ plane layer (top frames) and a CuO chain layer (bottom frames)
for three different interlayer interaction strengths (left, middle and
right frames). As the strength of the interlayer interaction,
$g_{12}$, is increased (left to right frames) the proportion of the
$s$-component in both layers increases. If the interlayer interaction
were further increased the gap nodes would leave the Brillouin zone
altogether and the order parameter would become $s$-like. Note that
the Fermi surface in the CuO$_2$ plane layer is tetragonal but that
the gap node is orthorhombic.  }
\label{fig11.ps}
\end{boxit}
\end{figure}
%\end{wrapfigure}

%\begin{wrapfigure}{t}{\columnwidth}
\begin{figure}
\begin{boxit}
	\begin{center} \begin{tabular}{c c c c}
		\makebox[0.22\columnwidth][l]{\large (a)} &
		\makebox[0.22\columnwidth][l]{\large (b)} &
		\makebox[0.22\columnwidth][l]{\large (c)} &
		\makebox[0.22\columnwidth][l]{\large (d)} \\
		\epsfxsize=0.22\columnwidth \epsfbox{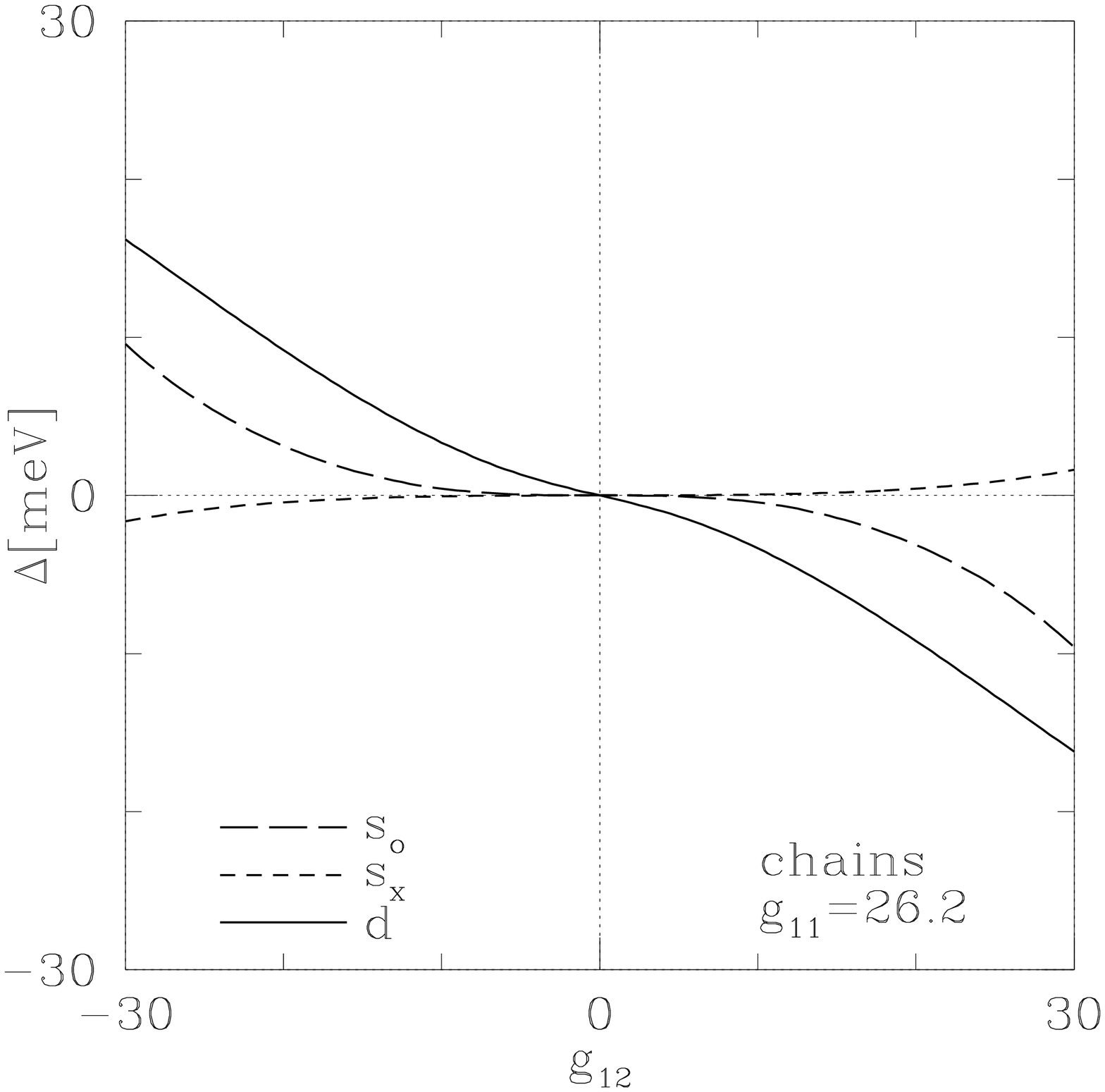} &
		\epsfxsize=0.22\columnwidth \epsfbox{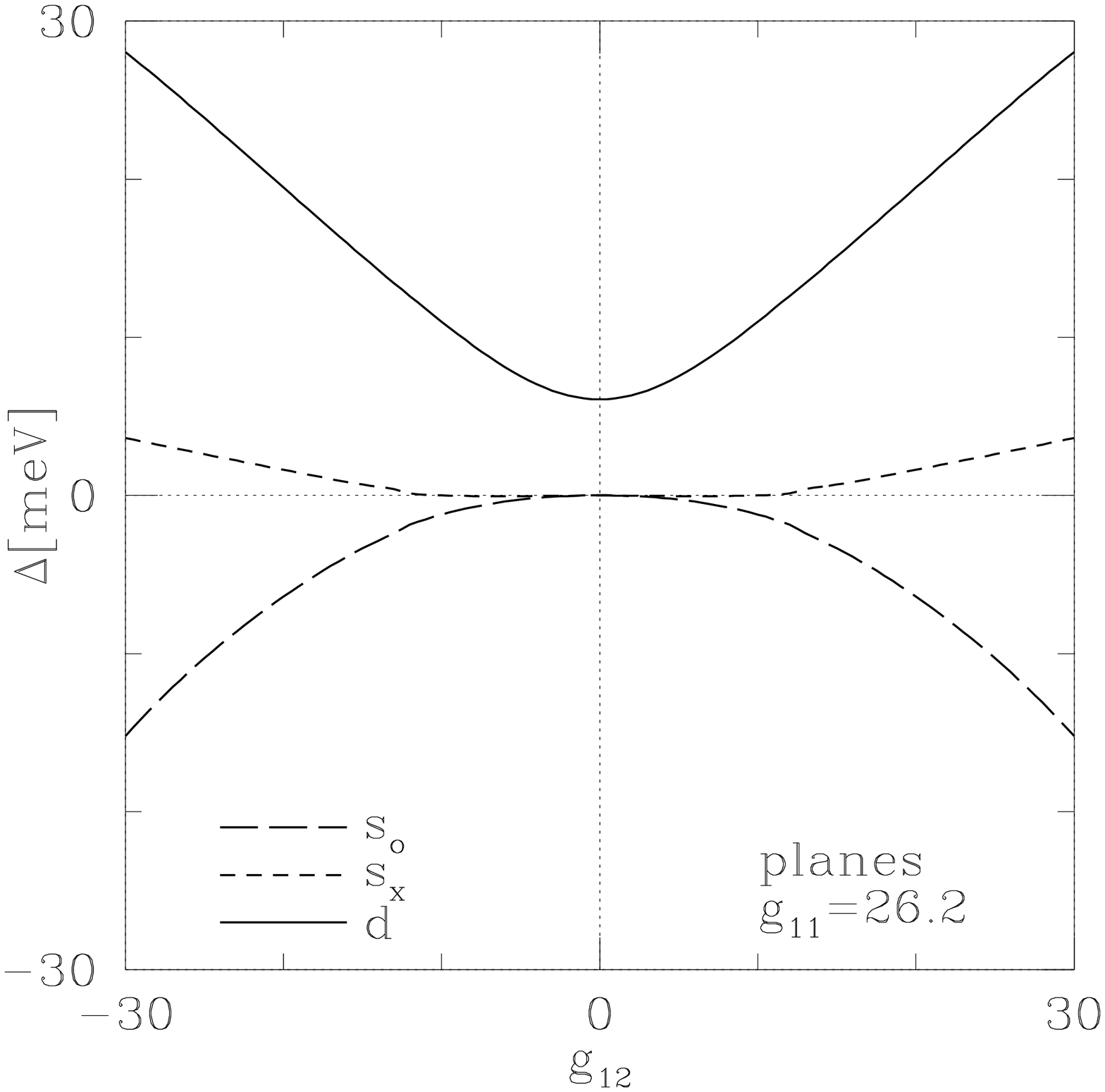} &
		\epsfxsize=0.22\columnwidth \epsfbox{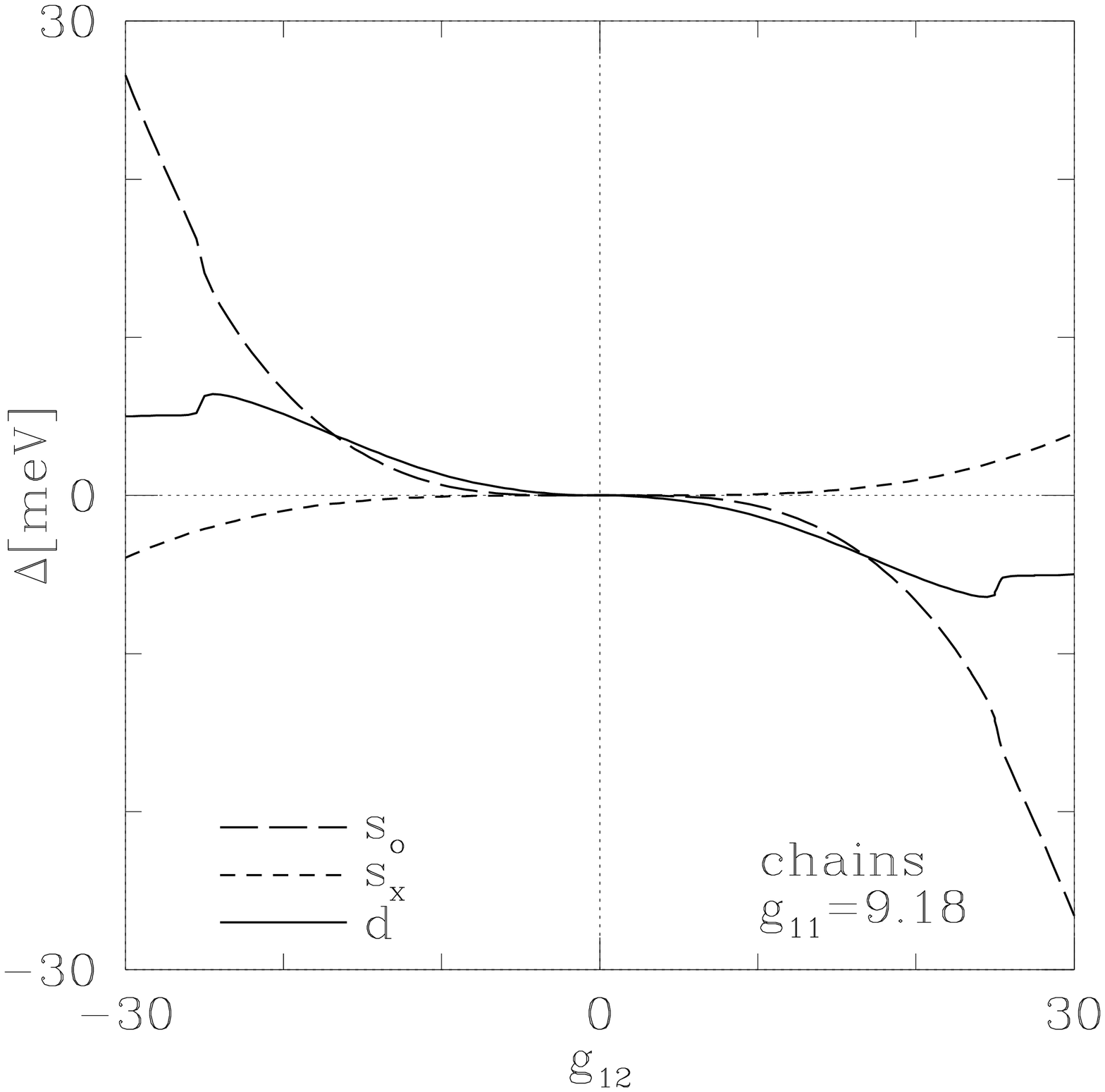} &
		\epsfxsize=0.22\columnwidth \epsfbox{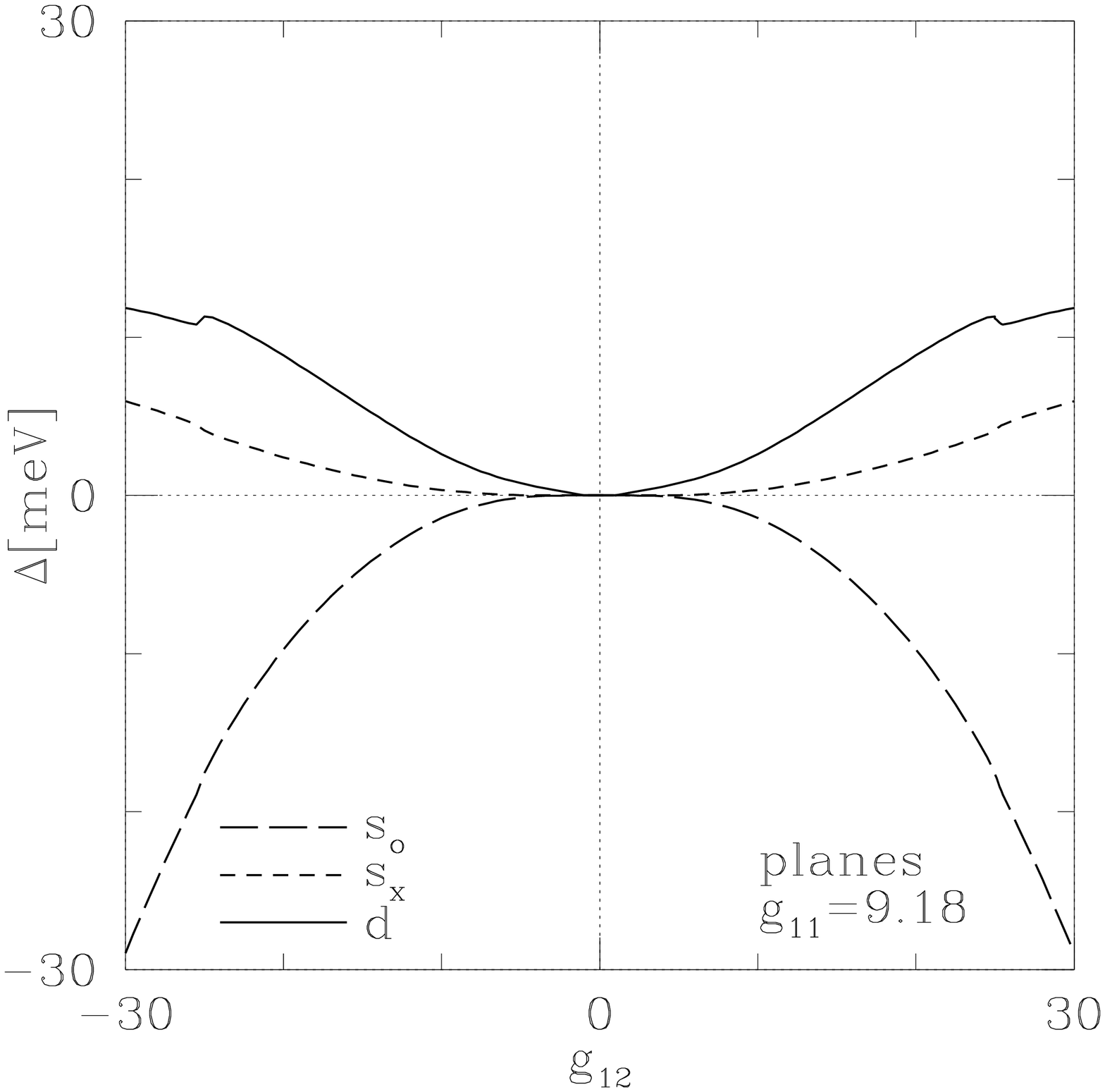}
	\end{tabular} \end{center}
\caption{\small 
Calculation of the zero temperature order parameters as a function of
the interlayer interaction, $g_{12}$, for two fixed values of the
interlayer interaction, $g_{11}$, (left and right pairs of frames)
presented for the planes, (b) and (d), and chains, (a) and (c),
separately. In all frames the solid curve is the $d$-wave component of
the order parameter, the short dashed curve is the extended $s$-wave
component and the long dashed curve is the isotropic $s$-wave
component. In the left frames $g_{11}=26.2$ and for $g_{12}=10$,
$T_{\rm c}=100K$. At $g_{12}=0$ the order parameter is zero in the
chains and is pure $d$-wave in the planes. As the interlayer
interaction is increased the order parameter becomes present in the
chains and there is a mixing of $s$-wave components. In the right
frames $g_{11}=9.18$ and for $g_{12}=20$, $T_{\rm c}=100K$. At
$g_{12}=0$ the order parameter is zero in both the chains and the
planes. As the interlayer interaction is increased the order parameter
becomes present in both the chains and planes and there is a mixing of
$s$-wave components with the isotropic $s$-wave component eventually
becoming dominant. The feature at $g_{12}\sim 25$ occurs when the gap
node leaves the Brillouin zone.  As discussed in the text there is a
$g_{12}\leftrightarrow -g_{12}$ symmetry. Both $g_{11}$ and $g_{12}$
are in units of $t_1$; $g_{22}$, the coupling in the chains, is set
equal to zero.}
\label{fig12.ps}
\end{boxit}
\end{figure}
%\end{wrapfigure}

If one wants to remain within a model where there is no
pairing interaction in the chains, one way to increase the value of
the chain gap is to include off diagonal pairing in a two band {\sc bcs}
model.  The {\sc bcs} equations in this case are [18-19]
\begin{eqnarray}
\label{bcs.eq}
\mit\Delta_{{\bf k},1}&=& \frac{1}{\Omega}\sum_{{\bf q}}{\left(
V_{{\bf k},{\bf q},11}\chi_{{\bf q},1}
+V_{{\bf k},{\bf q},12}\chi_{{\bf q},2}
\right)} \nonumber \\
\mit\Delta_{{\bf k},2}&=& \frac{1}{\Omega}\sum_{{\bf q}}{\left(
V_{{\bf k},{\bf q},12}\chi_{{\bf q},1}
+ V_{{\bf k},{\bf q},22}\chi_{{\bf q},2}
\right)},
\end{eqnarray}
where the pairing susceptibility is:
\begin{eqnarray}
\label{sus.eq}
\chi_{{\bf q},\alpha}&\equiv&
\left<c_{{\bf q}\uparrow,\alpha} c_{-{\bf q}\downarrow,\alpha}\right> \nonumber \\
&=&\frac{\mit\Delta_{{\bf q},\alpha}}{2E_{{\bf q},\alpha}}
\tanh \left( \frac{E_{{\bf q},\alpha}}{2 k_{\rm B}T} \right),
\end{eqnarray}
for the $\alpha$'th band with the brackets $\left< \cdot \right>$
indicating a thermal average of the pair annihilation operators.  In
calculations, we will assume $V_{22}$ to be zero but take $V_{12} =
V_{21}$ finite.  This is the term that couples the chains and planes
and makes the chains have a gap.  For the electron dispersions, we
take a model with up to second neighbour hopping with:
\begin{eqnarray}
\label{disp.eq}
\xi_{{\bf k},\alpha}&=&
-2t_\alpha\left[(1+\epsilon_\alpha)\cos(k_x)+\cos(k_y)\right.\nonumber\\
&&-\left.2B_\alpha\cos(k_x)\cos(k_y)-(2-2B_\alpha-\mu_\alpha)\right],
\end{eqnarray}
where the two new parameters not in equation (1) and (2) are the
second neighbour hopping $B_\alpha$ and the orthorhombic distortion
$\epsilon_\alpha$. As a model we take $\{t_\alpha, \epsilon_\alpha, B_\alpha,
\mu_\alpha\}$ to be $\{100, 0, 0.45, 0.51\}$ for the planes and $\{-50,
-0.9, 0, 1.2\}$ for the chains.  The resulting Fermi surfaces for
chains (long dashes) and planes (short dashes) are shown in Fig.~9.

To solve for the gaps of equations (\ref{bcs.eq}), we need to make
some choices for the pairing potential $V_{{\bf k},{\bf
k^\prime},\alpha\beta}$.  In the previous section, we chose a
separable form.  Here we use a different alternative, which is based
on the nearly antiferromagnetic Fermi liquid approach, and assume:
\begin{eqnarray}
V_{{\bf k},{\bf q},\alpha\beta}=g_{\alpha\beta}
\frac{-t}{1+\xi_\circ^2|{\bf k-q-Q}|^2},
\label{mmp.eq}
\end{eqnarray}
which is proportional to the antiferromagnetic spin susceptibility
with magnetic coherence length $\xi$ taken from Millis, Monien and
Pines \cite{mmp} and $t=100$meV sets the energy scale. In equation
(\ref{mmp.eq}), $\bf Q$ is the commensurate wave vector $(\pi, \pi)$ and
therefore the repulsive interaction (\ref{mmp.eq}) is peaked at $(\pi,
\pi)$.  This interaction leads directly to a $d_{x^2-y^2}$ gap in a
single plane.  If the coupling $g_{\alpha\beta}$ for $\alpha\not
=\beta$ is different from zero superconductivity is induced in the
chains and the gap no longer has pure $d_{x^2-y^2}$ symmetry in both
chains and planes.  It will be an admixture of:
\begin{eqnarray*}
\eta_{\bf k}^{(s_\circ)}&=& 1\nonumber \\
\eta_{\bf k}^{(s_{x^2+y^2})}&=& \cos(k_x)+\cos(k_y)\nonumber \\
\eta_{\bf k}^{(d_{x^2-y^2})}&=& \cos(k_x)-\cos(k_y).
\label{basis.eq}
\end{eqnarray*}

In Fig.~10, we show results for a case $\{g_{11}, g_{12}, g_{22}\}=
\{26.2, 10, 0\}$ with the critical temperature taken to be 100 K
representative of the copper oxides.  What is shown in frames (a) and
(b) are the gaps as a function of momentum in the first Brillouin zone
for the plane and chain, respectively.  The projection on d-wave and
s-wave manifold are also shown in (c) and (e) for the plane and (d)
and (f) for the chain.  We see that the orthorhombic chains can lead
to a significant mixture of $s$ and $d$ components even for the plane
case.

       A useful representation of these gap results is to show the
contours of gap zeros on the same plot as the Fermi surface.  This is
presented in the series of frames shown in Fig.~\ref{fig11.ps}.  The
top frames apply to the planes while the bottom frames apply to the
chains.  In all cases, (a), (c), (e) for the planes and (b), (d), (f)
for the chains, the same Fermi surface (dashed curves) was used.  By
choice, the Fermi contour have tetragonal symmetry in the top figure
while the chain Fermi surface is a quasi straight line along $k_x$ as
is expected for chains along $y$ in configuration space.  The pictures
are for three different values of pairing potential.  The first set of
two frames (a) and (b) are for $\{g_{11}, g_{12}, g_{22}\} =
\{29.9, 5, 0\}$, i.e. very little coupling between chains and planes
(off diagonal $g_{12}$ small).  In this case, the gap in the plane is
nearly pure d-wave as is also the induced gap in the chains.  As the
coupling $g_{12}$ is increased to $\{g_{11}, g_{12}, g_{22}\} = \{26.2,
10, 0\}$ a significant s-wave component gets mixed into both solutions
and the gap nodes move off the main diagonals of the Brillouin zone.
(This is the solution that is plotted in Fig.~10).  The gap nodes
still cross the Fermi surfaces in both chains and planes.  As the
coupling is further increased to $\{g_{11}, g_{12}, g_{22}\} = \{9.18,
20, 0\}$, Fig.~11 (e) and (d), the gap nodes move far off the diagonal
and for the chains they no longer cross the Fermi surface so that
there is a finite minimum value of the gap on this sheet of the Fermi
surface.

       The amount of admixture of each of the three components in
equation (\ref{basis.eq}) are shown as a function of off diagonal
$g_{12}$ in Fig.~12 for the last two cases, namely $g_{22} = 26.2$
left frames and $g_{11} = 9.18$ right frames.  The amplitude of the
gap component involved is given in meV.  Frames (a) and (c) are for
the chains while frames (b) and (d) are for the planes.  In all cases,
the solid curve is the $d_{x^2-y^2}$-component, short dashed the
$s_{x^2+y^2}$-component and long dashed the $s_\circ$-component.

For the first choice of intralayer interaction (left frames),
$g_{11}=26.2$, and there is no order parameter in the chains when
there is no interlayer interaction (ie, $g_{12}=0$) and the order
parameter in the planes is pure $d$-wave. As the interlayer
interaction is increased from zero, $s$-wave components appear in the
planes and all three components appear in the chains. This ``$s+d$
mixing'' is caused by the breaking of the tetragonal symmetry upon the
introduction of the chains; there is no relative phase between the
$s$- and $d$-wave components within either the planes or chains but
there can be a relative phase between the order parameter in the
planes and chains. In the range of $g_{12}$ explored here the $d$-wave
component in the plane remains dominant but for sufficiently strong
interlayer interaction the isotropic $s$-wave component eventually
dominates~\cite{liechtenstein} (ie, the gap nodes disappear).  For
interaction parameters $\{g_{11},g_{12}, g_{22}\} =
\{26.2,10,0\}$ the critical temperature is 100K and the maximum value
of the gap in the Brillouin zone is 27.5meV in the planes and 8.0meV
in the chains, while the maximum values on the Fermi surfaces are
approximately 22meV and 7meV respectively. The ratio $2\mit\Delta_{\rm
max}/T_{\rm c}$ is 6.4 in the planes and 1.9 in the chains.

For the second choice of intralayer interaction (right frames),
$g_{11}=9.18$, there is no order parameter in either the chains or the
planes when there is no interlayer interaction (ie, $g_{12}=0$). As
the interlayer interaction is increased $d$-wave and then $s$-wave
components of the order parameter appear in both the planes and
chains. Again, there is no relative phase between the $s$- and
$d$-wave components within either the planes or chains but there can
be an overall relative phase between the order parameter in the planes
and chains. At approximately $g_{12}=15$ the gap nodes no longer cross
the Fermi surface in the chains; the feature at $g_{12}\sim 25$
coincides with the gap nodes leaving the Brillouin zone and isotropic
$s$-wave becoming dominant.  For interaction parameters
$\{g_{11},g_{12}, g_{22}\} = \{9.18,20,0\}$ the critical temperature
is again 100K and the maximum value of the gap in the Brillouin zone
is now 32.8meV in the planes and 20.1meV in the chains, while the
maximum values on the Fermi surfaces are approximately 27meV and 17meV
respectively. The ratio $2\mit\Delta_{\rm max}/T_{\rm c}$ is 7.6 in
the planes and 4.7 in the chains.

Note that for $g_{12}>0$ all of the $s$-wave components of the order
parameters in both the planes and chains have the same relative sign
and the $d$-wave components have opposite signs, while for $g_{12}<0$
all of the relative signs are reversed but that the magnitudes of the
components are insensitive to the sign of $g_{12}$.

\myfigure{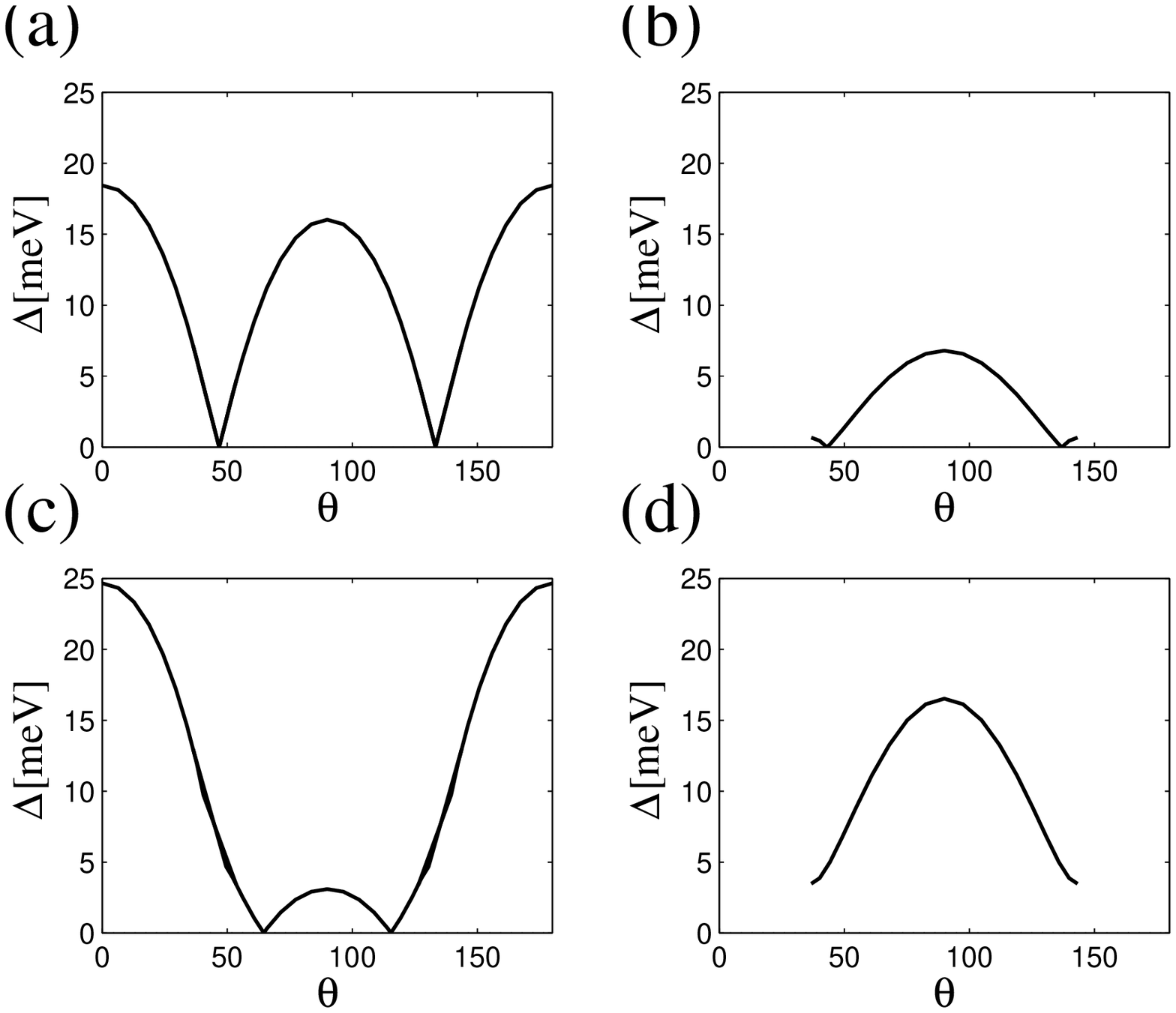}{0.6}
{The magnitude of the gap on the Fermi surface as a function of angle
for the four cases of Figs.~\ref{fig12.ps}. The angle $\theta$ is
measured from the center or $(\pi,\pi)$ point of the Brillouin zone
with the $y$-axis (ie, the vertical) corresponding to
$\theta=0$. Frames (b) and (d) do not span all angles due to the Fermi
surface not being closed in the chain layer. For the first choice of
interaction parameters, $\{g_{11},g_{12},g_{22}\} = \{26.2,10,0\}$,
the ratio $2\mit\Delta_{\rm max(FS)}/T_{\rm c}$, where
$\mit\Delta_{\rm max(FS)}$ is the maximum value of the gap on the
Fermi surface, is 4.3 and 1.6 for the planes and chain respectively;
for the second, $\{g_{11},g_{12},g_{22}\} =\{9.18,20,0\}$, they are
5.7 and 3.8.}

In Fig.~\ref{fig13.ps} the magnitude of the gap is plotted as a function
of angle, $\theta$, along the Fermi surface measured from the
vertical. In frame (a) of Fig.~\ref{fig13.ps} the local maxima of
the gap on the Fermi surface are 16 and 18meV; in (b) they are 1 and
7meV, and in (c) they are 25 and 3meV. In (d) one can see that there
are no gap nodes which cross the Fermi surface; the maximum and
minimum value of the gap on the Fermi surface are 17 and 4meV
respectively.

\begin{figure}
%\begin{wrapfigure}{t}{0.65\columnwidth}
\begin{boxit}
	\begin{center}
		\begin{tabular}{c c}
			\makebox[0.25\columnwidth][l]{\large (a)} &
			\makebox[0.25\columnwidth][l]{\large (b)} \\
			\epsfxsize=0.25\columnwidth
			\epsfbox{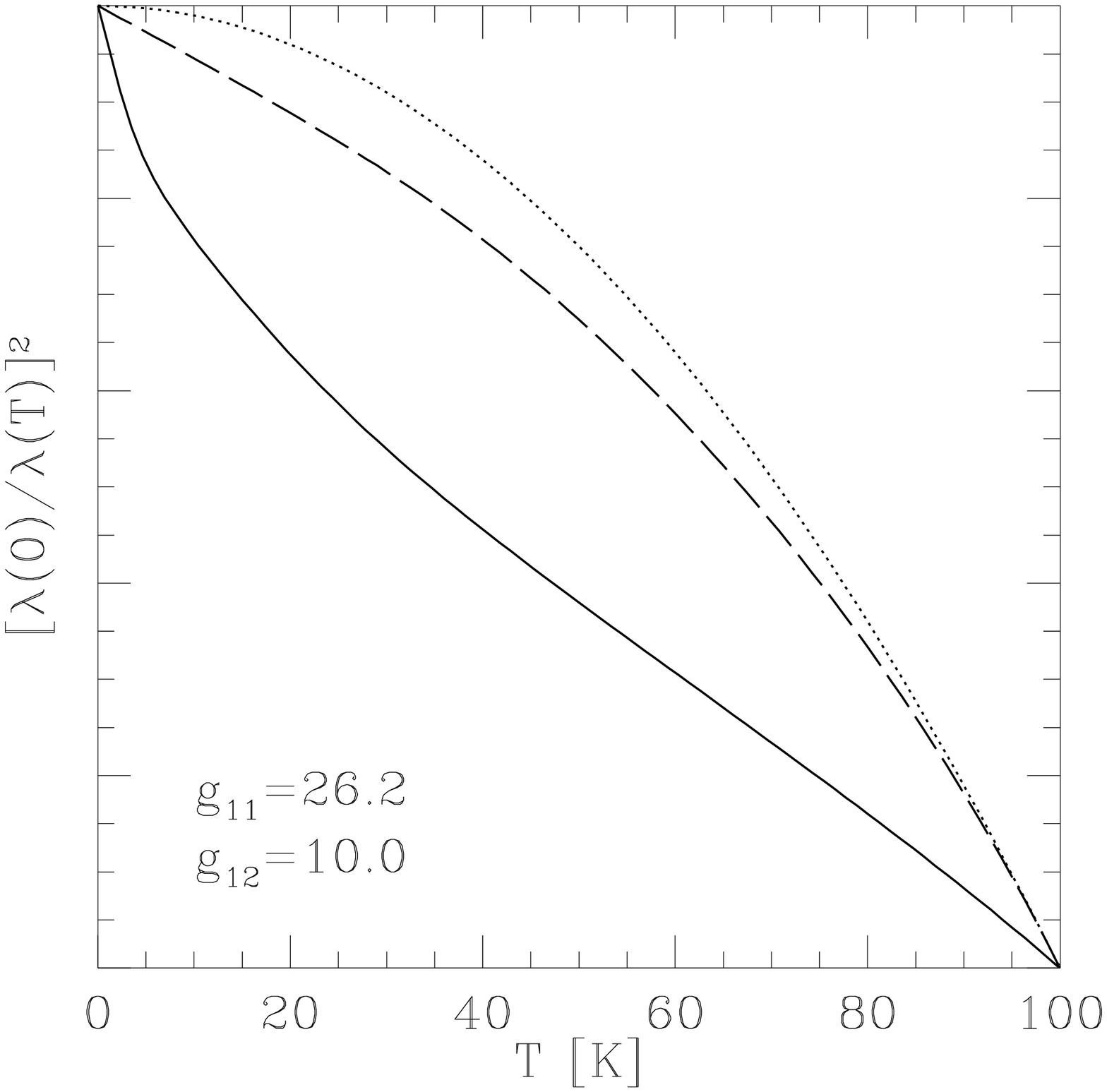} &
			\epsfxsize=0.25\columnwidth
			\epsfbox{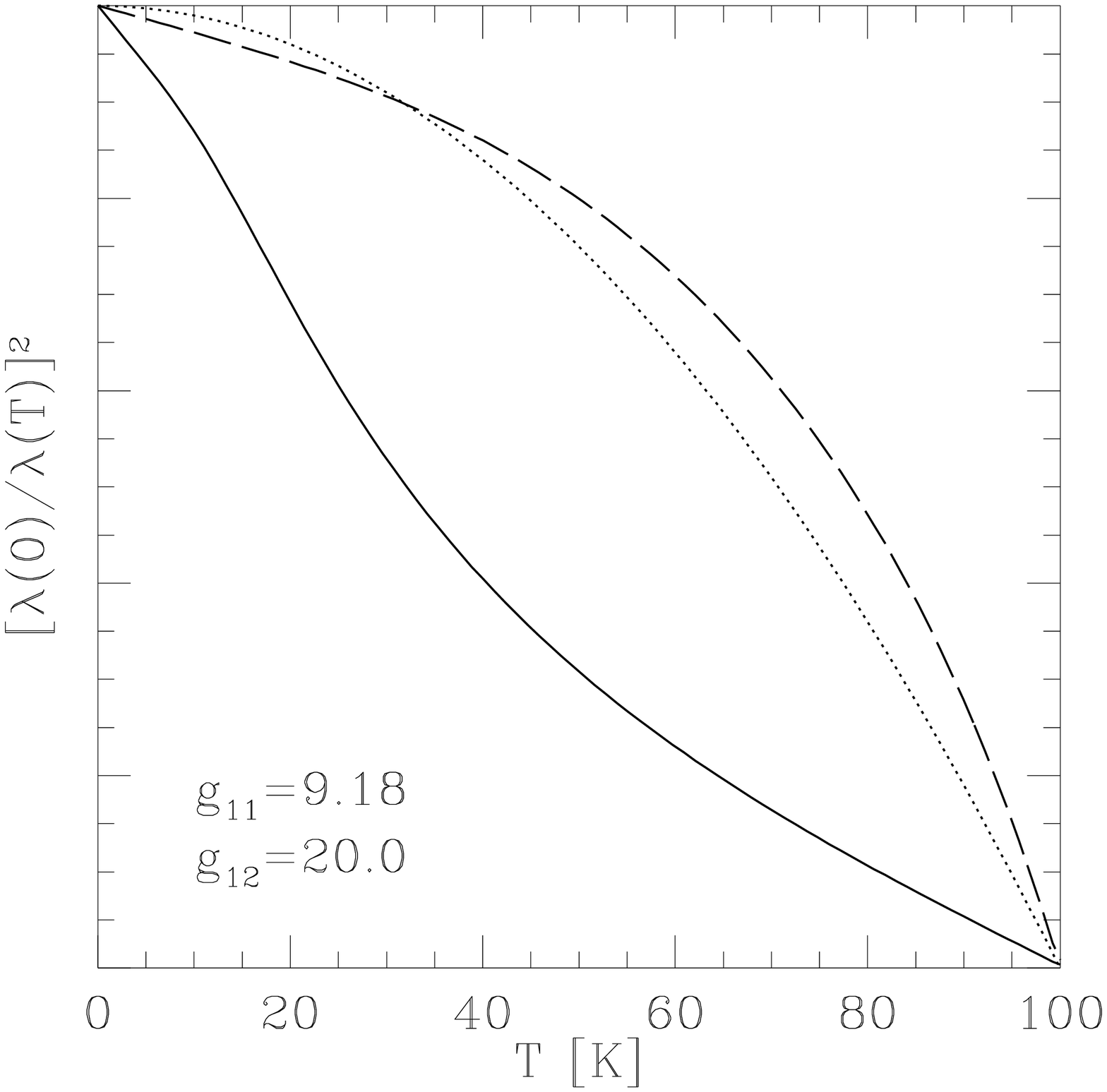} \\ 
		\end{tabular}
	\end{center}
\caption{\small
The magnetic penetration depth for the two sets of interaction
parameters.  Which show the magnetic penetration depth along (solid
curve) and perpendicular (dashed curve) to the CuO chains.  The dotted
curve is $1-(T/T_{\rm c})^2$ and is shown for comparison. The chains,
due to their Fermi surface, do not contribute appreciably to the
penetration depth perpendicular to the chains (dashed curves).  The
ratio $(\lambda_{yy}/\lambda_{xx})^2$ is 1.37 for both sets of
parameters since this is a normal state property. }
\label{fig14.ps}
\end{boxit}
\end{figure}
%\end{wrapfigure}

In Fig.~\ref{fig14.ps} we have plotted the magnetic penetration depth
calculated with the lowest three harmonics (\ref{basis.eq}) of the
solutions of the {\sc bcs} equations (\ref{bcs.eq}) for the two
choices of interaction parameters. The solid curve is for the
$y$-direction (along the chains) and the dashed curve is for the
$x$-direction (perpendicular to the chains). The dotted curve is
$1-(T/T_{\rm c})^2$ and is plotted for comparison. The ratio
$\lambda_{yy}/\lambda_{xx}$ at zero temperature is 1.37 for both
interaction parameter choices since the zero temperature penetration
depth is a normal state property. The zero temperature penetration
depth is largely governed by the bandwidth (ie,
$4t_\alpha(2-\epsilon_\alpha)$) -- the larger the bandwidth the larger
the zero temperature penetration depth.

As pointed out above, the curvature of the penetration depth curve,
$\lambda_{ii}^{-2}(T)$, is largely governed by the ratio
$2\mit\Delta_{\rm max}/T_{\rm c}$ and is a straight line for the
$d$-wave {\sc bcs} value of $4.4$. The presence of the chain layer and
the interlayer interaction increases this ratio in the plane layer but
it remains low in the chain layer due to the absence of an interaction
in this layer. It is this lower value that makes
$\lambda^{-2}_{yy}(T)$ (along the chains) have upward curvature (solid
curves). Including pairing in the chains will push the solid curve
towards the dashed one and make the initial low temperature slopes
fall closer to each other.

\section*{Conclusions}

In conclusion, the present data on single crystal untwinned YBCO at
optimum doping suggest that the proximity effect incorporated into a
single perpendicular tunnelling parameter $t_\perp$ cannot account for
the observation.  If interplane pairing is included the situation is
greatly improved provided the off diagonal pairing is increased
sufficiently to produce a gap on the chain which is of the same order
of magnitude as that in the planes.  Similar values of the gaps on the
chains and planes can be taken as evidence that optimally doped YBCO
is fairly three dimensional and that the coherence length in the
$z$-direction may not be sufficiently short to allow spatial
inhomogeneities to exist within a unit cell.

%\newpage
\section*{Acknowledgements}

Research supported in part by the Natural Sciences and Engineering
Research Council of Canada ({\sc nserc}) and by the Canadian Institute
for Advanced Research ({\sc ciar}).

\end{document}